\documentclass[aps,prc,superscriptaddress, floatfix, twocolumn, nofootinbib]{revtex4-1}
\usepackage{graphicx,subfigure}
\usepackage{amsmath,amssymb,amsfonts}
\usepackage{bm}
\usepackage{hyperref}
\usepackage{color}
\usepackage{lineno}
\usepackage{soul}
\setulcolor{blue}
\setstcolor{red}
\sethlcolor{yellow}

\def \bes {\begin{subequations}}
\def \ees {\end{subequations}}

\def \be {\begin{equation}}
\def \ee{\end{equation}}

\def \pd {\partial}

\def \eq {Eq.~}

\def \a {\alpha}
\def \b {\beta}

\def \d {\delta}
\def \e {\epsilon}

\def \k {\kappa}
\def \o {\omega}

\def \l {\lambda}

\def \s {\sigma}

\def \<{\langle}
\def \>{\rangle}
\def \+{\dagger}
\def \({\left(}
\def \){\right)}
\def \[{\left[}
\def \]{\right]}

\def \equ {\text{eq}}

\def \eff{\text{eff}}

\def \IS {\text{Israel-Stewart~}}

\graphicspath{{/plots/}}

\begin{document}

\title{Phenomenological Consequences of Enhanced Bulk Viscosity Near the QCD
Critical Point}

\author{Akihiko Monnai}
\affiliation{RIKEN BNL Research Center, Brookhaven National Laboratory, Upton, NY
11973, USA }
\affiliation{Institut de Physique Th\'{e}orique, CNRS URA 2306, CEA/Saclay, F-91191
Gif-sur-Yvette, France}

\author{Swagato Mukherjee}
\affiliation{Department of Physics, Brookhaven National Laboratory, Upton, New York
11973-5000}

\author{Yi Yin}
\affiliation{Department of Physics, Brookhaven National Laboratory, Upton, New York
11973-5000}
\affiliation{Center for Theoretical Physics, Massachusetts Institute of Technology, Cambridge, MA 02139 USA}

\date{ \today}


\begin{abstract}

In the proximity of the QCD critical point the bulk viscosity of quark-gluon matter
is expected to be proportional to nearly the third power of the critical correlation
length, and become significantly enhanced.  This work is the first attempt to study
the phenomenological consequences of enhanced bulk viscosity near the QCD critical
point.  For this purpose, we implement the expected critical behavior of the bulk
viscosity within a non-boost-invariant, longitudinally expanding $1+1$ dimensional
causal relativistic hydrodynamical evolution at non-zero baryon density. We
demonstrate that the critically-enhanced bulk viscosity induces a substantial
non-equilibrium pressure, effectively softening the equation of state, and leads to
sizable effects in the flow velocity and single particle distributions at the
freeze-out. The observable effects that may arise due to the enhanced bulk viscosity
in the vicinity of the QCD critical point can be used as complimentary information to
facilitate searches for the QCD critical point. 

\end{abstract}

\maketitle

\section{Introduction}

The structure of the QCD phase diagram has attracted much attention and triggered a
plethora of theoretical and experimental studies (see
Refs.~\cite{Stephanov:2004wx,Stephanov:2007fk,Fukushima:2010bq,Ding:2015ona,STAR-wp,Heinz:2015tua}
for reviews). Of high interest is the existence of a conjectured critical
point~\cite{Asakawa:1989bq,Barducci:1989wi,Halasz:1998qr,Berges:1998rc,Stephanov:1998dy}
in the QCD phase diagram.  This critical point is the end point of the first-order
phase transition line that separates, in the chiral limit, a chirally symmetric
quark-gluon plasma (QGP) phase from the hadron-matter phase of QCD. The existence of
this critical point has been supported by many models of QCD thermodynamics.
However, its precise location in the phase diagram and even its existence, is
uncertain from the first-principle lattice simulations~\cite{Ding:2015ona}. 
  
An entire experimental program, the Beam Energy Scan (BES) at the Relativistic
Heavy-Ion Collider (RHIC) aims to search for the QCD critical
point~\cite{STAR-wp,Heinz:2015tua}. A universal feature of a system near a critical
point is the emergence of a critical mode, with growing and eventually divergent
correlation length, $\xi$. Thus, physical quantities which are more sensitive to the
growth of critical correlation length are expected to play crucial roles in
experimental searches for the QCD critical point. 

Well-known and more extensively studied examples are the non-Gaussian fluctuations of
the critical mode which grow as high powers of $\xi$~\cite{Stephanov:2008qz}. For
example, while the variance grows as $\xi^2$, the skewness and the kurtosis are
expected to grow more rapidly as $\xi^{4.5}$ and $\xi^7$, respectively. These
enhanced near critical fluctuations are accessible through measurements of
event-by-event fluctuations of particle
multiplicities~\cite{Stephanov:1998dy,Hatta:2003wn,
Stephanov:2008qz,Athanasiou:2010kw}. Expected growth of these fluctuation measures
reflect the static properties of the underlying near critical background. 

In the present work we pursue a complementary avenue --- we focus on how the growth
of $\xi$ near the QCD critical point affects the bulk hydrodynamic evolution of the
medium, and explore potential observables sensitive to such critical dynamics.

Hydrodynamical transport coefficients of a system near criticality scale with the
correlation length with universal exponents, which are fixed by the dynamical
universality class of the system. The dynamical universality class of the QCD
critical point is argued to be that of
model-H~\cite{Son:2004iv,Fujii:2003bz,Fujii:2004za,Fujii:2004jt} according to the
classification of Ref.~\cite{RevModPhys.49.435}. For a system belonging to the
dynamical universality class of model-H~\footnote{
Throughout this work we use approximate rational values of critical exponents: 
$\left(\alpha, \beta, \gamma, \delta, \nu, \eta \right)=\left(0, 1/3,4/3,5,2/3, 0 \right)$. 
These approximate values are within few percents of their respective exact values.
}
\begin{equation}
\label{transport-other}
\eta \sim  \xi^{\frac{1}{19}\epsilon}\, , 
\qquad
\l \sim \xi\, , 
\qquad \mathrm{and} \qquad
D_{B}\sim 1/\xi \, ,
\end{equation}
where $\eta$, $\l$, and $D_{B}$ denote the shear viscosity, thermal conductivity, and
baryon diffusion constant, respectively. $\epsilon=4-d$, where $d$ is the spatial
dimension. 
More importantly, near the criticality the bulk viscosity is expected to
grow far more rapidly~\cite{Moore:2008ws,PhysRevE.55.403} (see also
Sec.~\ref{sec:bulk_CEP})
\begin{equation}
\label{zeta_CEP}
\zeta \sim \xi^{3}\, . 
\end{equation}
In view of the above behaviors of the transport coefficients, it is crucial to
understand how the bulk hydrodynamical evolution of the matter created in heavy-ion
collisions will be modified in the proximity of the QCD critical point, and what are
the possible phenomenological consequences of such modification, if any.  Previously,
the hydrodynamical evolution near a critical point has been studied in a number of
references~\cite{Paech:2003fe,Nonaka:2004pg,Paech:2005cx,Herold:2013bi} (see
Ref.~\cite{Nahrgang:2016ayr} for a recent review). 
While the critical behavior of the baryon diffusion constant was addressed in 
Ref.~\cite{Kapusta:2012zb}, studies incorporating critical behavior of bulk viscosity 
are sorely lacking.

This work is the first attempt to address this issue.  We focus on the critical
behavior of bulk viscosity as it exhibits the strongest dependence on the correlation
length among the transport coefficients. As the first attempt, we incorporate the
critical behavior of the bulk viscosity (\emph{c.f.} Eq.~\eqref{zeta_CEP}) within a
non-boost-invariant, longitudinally expanding $1+1$ dimensional causal relativistic
hydrodynamical equations (\emph{i.e.} the \IS theory \cite{Israel:1979wp}) at non-zero baryon density.
Along the way, we also discuss the behavior of the bulk relaxation time,
$\tau_{\Pi}$, in the vicinity of the critical point. $\tau_{\Pi}$ is a transport
coefficient in \IS theory that controls the relaxation time of bulk viscous pressure
towards its Navier-Stokes limit. We argue that near the QCD critical point
$\tau_{\Pi}\sim \xi^{3}$. To best of our knowledge, the critical behavior of
$\tau_{\Pi}$ has not been discussed before. 

The rest of the paper is organized as follows. In Sec.~\ref{sec:bulk_CEP}, we review
the behavior of bulk viscosity and discuss the behavior of bulk relaxation time near
the QCD critical point.  We explain our set-up for hydrodynamic evolution in
Sec.~\ref{sec:set_up}. We present our results in Sec.~\ref{sec:results} and summarize
in Sec.~\ref{sec:summary}.

\section{Bulk viscosity and relaxation time near QCD critical point}
\label{sec:bulk_CEP}

We consider the following second order viscous hydrodynamic equations with finite
baryon density
\bes
\begin{eqnarray}
\label{Tmunu}
\nabla_\mu T^{\mu \nu} &=& 0 \, ,\\
 \label{IS}
\nabla_{\mu}J^{\mu}_{B} &=&0 \, , \\
 \label{Pi_rel} 
u^\mu \partial_\mu \Pi &=& -\frac{1}{\tau_\Pi} [\, \zeta \partial_\mu u^\mu+\Pi\, ]
\, ,
\end{eqnarray}
\label{IS-eq}
\ees
where $\nabla$ denotes the covariant derivative in Bjorken coordinates.  The
stress-energy tensor is decomposed as $T^{\mu\nu}=\e u^{\mu}u^{\nu} + \( p +\Pi \)
\Delta^{\mu\nu} +\pi^{\mu\nu}$, where $\e, p, u^{\mu}$ are the equilibrium energy
density, the equilibrium pressure and the fluid velocity, respectively.  The bulk
viscous pressure $\Pi$ measures the deviation of the pressure from its equilibrium
value.  $\Delta^{\mu\nu}=g^{\mu\nu}+u^{\mu}u^{\nu}$ projects onto the spatial
components in the local rest frame, and $\pi^{\mu\nu}$ is the shear-viscous stress
tensor. Our sign convention for the metric is chosen to be $g^{\mu
\nu}=\mathrm{diag}(-1,+1,+1,+1)$. Since the bulk viscosity exhibits the strongest
dependence on $\xi$, we will concentrate solely on the effects of the bulk viscosity.
More specifically, in Eq.~\eqref{IS-eq} we only keep terms involving the bulk
viscosity, $\zeta$, and the bulk relaxation time, $\tau_{\Pi}$, completely omitting
terms involving the shear viscous tensor, $\pi^{\mu\nu}$, as well as the baryon
density diffusion. More specifically, we use the baryon current $J^{\mu}_{B}=n_{B}
u^{\mu}$, $n_{B}$ being the baryon density, and do not include terms in the baryon
current which are proportional to the baryon diffusion.

Now, we discuss the behaviors of $\zeta$ and $\tau_{\Pi}$ near the QCD critical
point.  The behavior of $\zeta$ can be determined by noting that the dynamical
universal behavior of the QCD critical point is the same as for the liquid-gas phase
transition~\cite{Son:2004iv,Fujii:2003bz,*Fujii:2004za}, \emph{i.e.} that of
model-H~\cite{RevModPhys.49.435}.  Onuki has performed a detailed study of the
behavior of the bulk viscosity near the critical point of a liquid-gas
system~\cite{PhysRevE.55.403}, and it can be directly adapted for the case of QCD
critical point: $\zeta \sim \xi^{z-\alpha/\nu}$.  Here, $\alpha$ and $\nu$ are
standard equilibrium critical exponents,  and $z$ is the dynamical critical exponent.
At the mean-field level $\a=0$, and $z=3$ for the model-H~\cite{PhysRevE.55.403},
giving: $\zeta \sim  \xi^{3}$. 

An intuitive explanation of the behavior of bulk viscosity in the vicinity of the
critical point goes as follows~\cite{Moore:2008ws} --- bulk viscosity controls the
relaxation time of the pressure towards its equilibrium value after a rapidly applied
small compression or expansion.  Near a critical point, the pressure will remain out
of equilibrium until the slow critical mode relaxes back to its equilibrium value,
and, thus, $\zeta\propto\tau_{\s}$, where $\tau_{\s}$ is the relaxation time for the
critical mode. Due to the critical slowing down $\tau_{\s}$ provides the longest time
scale, and grows as $\tau_{\s} \sim \xi^{z}$~\cite{RevModPhys.49.435}. Hence, in the
proximity of a critical point the bulk viscosity is expected to be enhanced as
$\zeta\sim\tau_{\s}\sim\xi^z$.

The above discussion also sheds light on the behavior of $\tau_{\Pi}$ near a critical
point. Imagine a homogeneous equilibrated system, described by the equilibrium values
of $\e$, $p$, and with $\Pi=0$ and $u^{\mu}=(1,0,0,0)$. Now, consider a small
homogeneous, but time-dependent perturbation of the bulk pressure, $\delta\Pi$.
Since such a perturbation has no spatial dependence, $\delta\Pi$ does not change the
energy-momentum conservation equation Eq.~\eqref{Tmunu}. Consequently, $\pd_\mu u^\mu
=0$, and Eq.~\eqref{Pi_rel} becomes $\pd_{t}\delta\Pi=-\delta\Pi/\tau_{\Pi}$,
implying that the characteristic damping-time of the off-equilibrium pressure, $\Pi$,
is given by $\tau_{\Pi}$. Near a critical point this time scale should be governed by
the relaxation time of the slowest mode, \emph{i.e.} that of the critical mode, given
by $\tau_\s\sim\xi^z$. Thus, near a critical point it is natural to expect:
$\tau_{\Pi}\sim\tau_{\sigma}\sim\xi^z$.

The above argument can be further supplemented and strengthened by invoking causality
of the \IS theory. For the linearized \IS theory, neglecting the contribution of shear
viscosity, the dispersion relation of the sound mode is given by 
\cite{Romatschke:2009im}
\begin{equation}
\label{cs_infty}
\lim_{k\to\infty}
\frac{d\o_{\rm sound}(k)}{dk}= \sqrt{c^2_{s}+\frac{\zeta}{\tau_{\Pi}\(\e+p\)}}\, ,
\end{equation}
where $c_s$ denotes the speed of sound. To maintain causality the sound speed cannot
exceed the speed of light, \emph{i.e.} the right hand side of the above equation must
remain less than $1$. In other words, to ensure causality $\zeta/\tau_{\Pi}\(\e+p\)$
must remain finite, and cannot grow with the diverging correlation length near a
critical point. Since near a critical point $(\e+p)$ remains non-zero but the
bulk viscosity diverges as $\zeta\sim\xi^z$, causality dictates that the bulk
relaxation time also must grow at least as rapidly as $\tau_\Pi\sim\xi^z$.     

Thus, near a critical point the bulk relaxation time is expected to diverge as
\begin{equation}
\tau_\Pi \sim \xi^z \, .
\label{tau_Pi}
\end{equation}
While Eq.~\eqref{tau_Pi} is a natural consequence of slow relaxation of the critical
modes, to best of our knowledge, the critical behavior of $\tau_\Pi$ has not been
discussed in literature before.

To summarize, $\zeta$ and $\tau_{\Pi}$ determines the relaxation of pressure towards
its equilibrium value in long time limit, and, near a critical point, this relaxation
process is governed by the relaxation of the slowest critical modes. In particular,
for the QCD critical point belonging to the dynamical universality class of model-H,
we, therefore, have
\begin{equation}
\label{zeta-all-relatioin}
\zeta \sim \tau_{\Pi}\sim \tau_{\sigma} \sim \xi^{3}\, . 
\end{equation}
With the critical behaviors $\zeta$ and $\tau_\Pi$ at hand, we are now ready to study
their influences on hydrodynamic evolution near the QCD critical point. 

\section{Set-up for hydrodynamic evolution}
\label{sec:set_up}

\begin{figure}[htb]
\center
\subfigure[]
{
\includegraphics[width=0.45\textwidth]{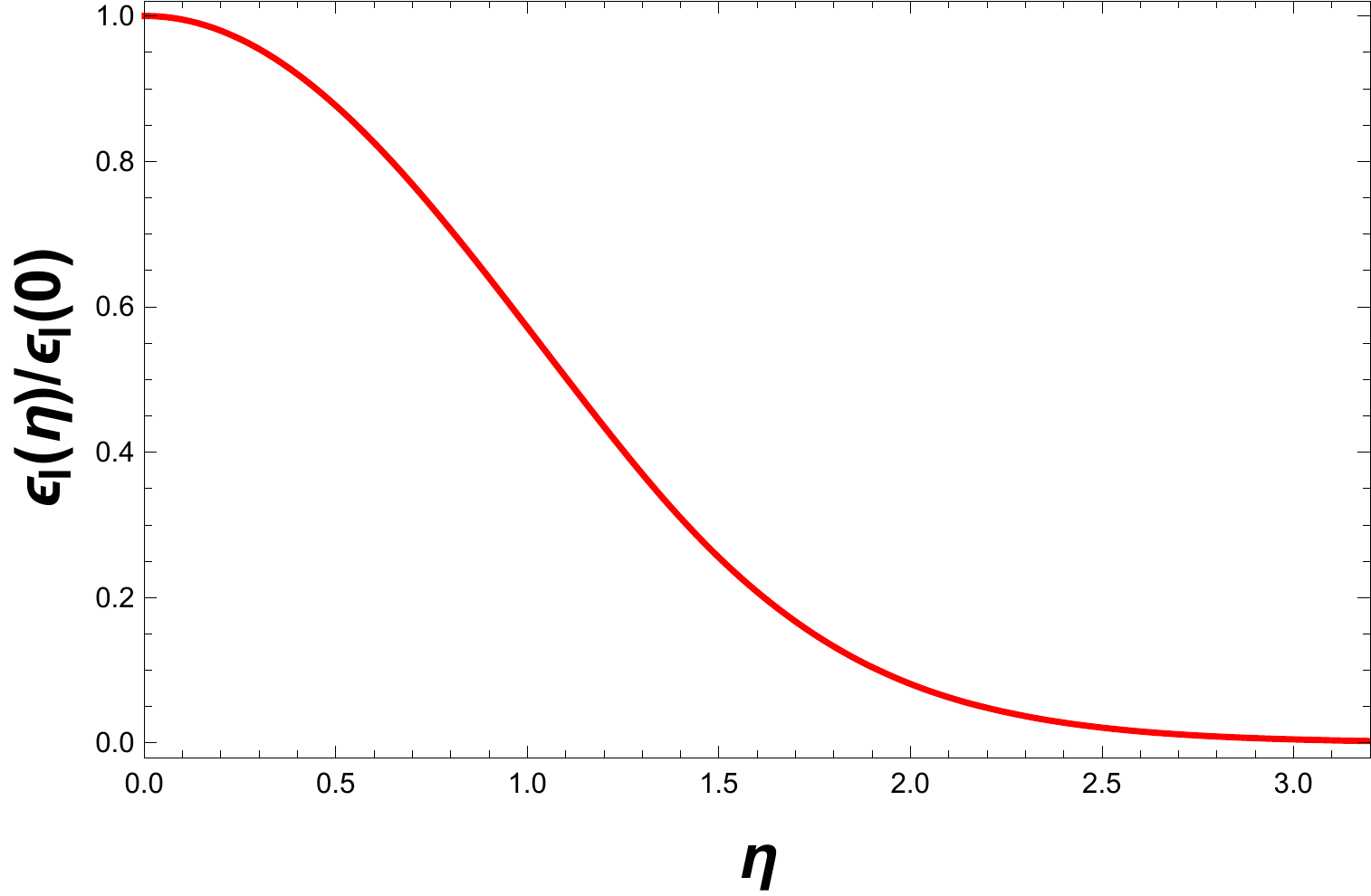}
\label{fig:EIC}
}
\subfigure[]
{
\includegraphics[width=0.45\textwidth]{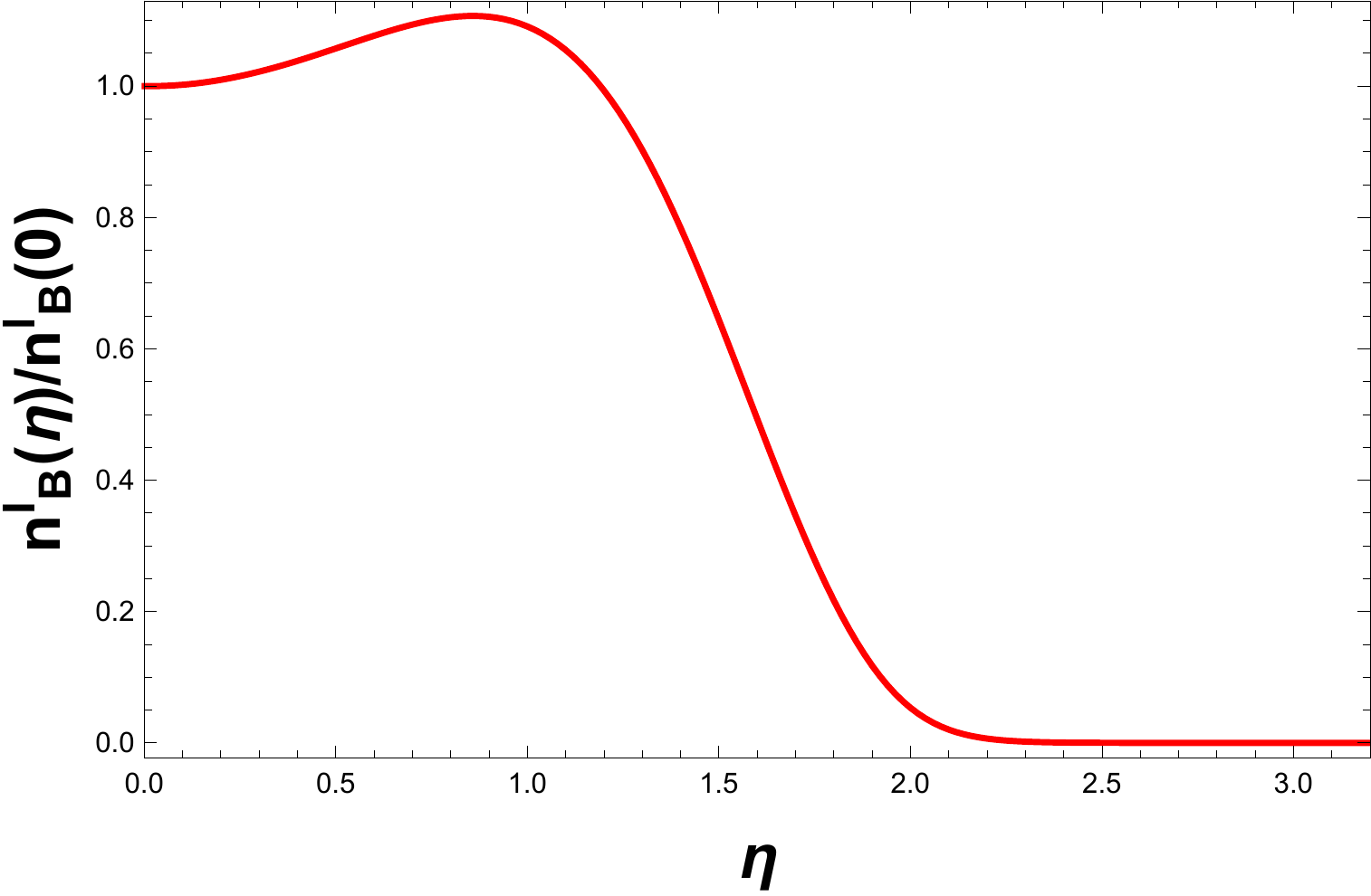}
\label{fig:NBIC}
}
\caption{(Color online) Dependence of initial energy density, $\epsilon_{I}(\eta)$ (a), and initial
baryon number density, $n^{I}_{B}(\eta)$ (b), normalized by their corresponding
values at $\eta=0$, on the spatial rapidity, $\eta$.  }
\label{fig:IC}
\end{figure}

In this exploratory study we consider $1+1$ dimensional \IS hydrodynamics with
longitudinal expansion along the $z$-direction \cite{Monnai:2012jc}. The temperature,
baryon density and fluid velocity only depend on the proper time
$\tau=\sqrt{t^{2}-z^{2}}$ and spatial rapidity $\eta=\arctan(z/t)$, and the fluid
velocity is given by $u^{\mu}=\( u^{\tau}, u^{\eta}, 0 ,0 \)$.  

We numerically solve the $1+1$ dimensional non-boost-invariant viscous hydrodynamic
equations, Eqs. \eqref{IS-eq}.  The spatial rapidity dependence of the initial energy
and baryon number densities are obtained by extrapolating color glass models
\cite{Drescher:2006ca, Drescher:2007ax, MehtarTani:2008qg, MehtarTani:2009dv} to the
fixed beam energy $\sqrt{s}=17$~GeV, see Fig.~\ref{fig:IC}. The normalizations and
parameters in those models are tuned to roughly imitate the rapidity distributions of the
charged hadrons and the net baryon number of the most central Pb-Pb collisions at SPS
near mid-rapidity at the corresponding beam energy. The initial state model used in
this study might not be completely realistic for heavy-ion collisions at the lower
energy. However, they suffice for this illustrative study. Since at this beam energy
the two incoming nuclei take longer to pass through each other, we start the
hydrodynamic evolution at $\tau_\mathrm{th} = 1.5$ fm/$c$. We further assume
$u^\eta(\tau_\mathrm{th})=0$ and $\Pi(\tau_\mathrm{th})=0$.

Hydrodynamical equations are closed by providing an equation-of-state (EoS),
$p(\mu_{B},T)$, $\e(\mu_{B},T)$ and $n_B(\mu_{B},T)$. The EoS is constructed out of
the second- and fourth-order net-baryon number fluctuations computed using lattice
QCD ~\cite{Bazavov:2014pvz, Bazavov:2012jq, Ding:2015fca}, and by matching onto that
obtained from a hadron resonance gas model at low temperatures \cite{Monnai:2015sca,
Denicol:2015nhu}. As our sole focus is on the effects of the critical enhancement of
bulk viscosity, for simplicity, we do not include any critical behavior in the EoS
itself.

For further quantification this critical enhancement, we consider two cases with the
same initial condition. In the first case, denoted by ``no-CP'', we solve
\eqref{IS-eq} by assuming that the QCD critical regime is far away from the evolution
trajectories in the $\mu_B$-$T$ plane.  In the second case, denoted by ``CP'', the location
and width of the critical region are chosen to make some evolution trajectories
pass through the critical regime.

\begin{figure}[htb]
\centering
\includegraphics[width=.48\textwidth]{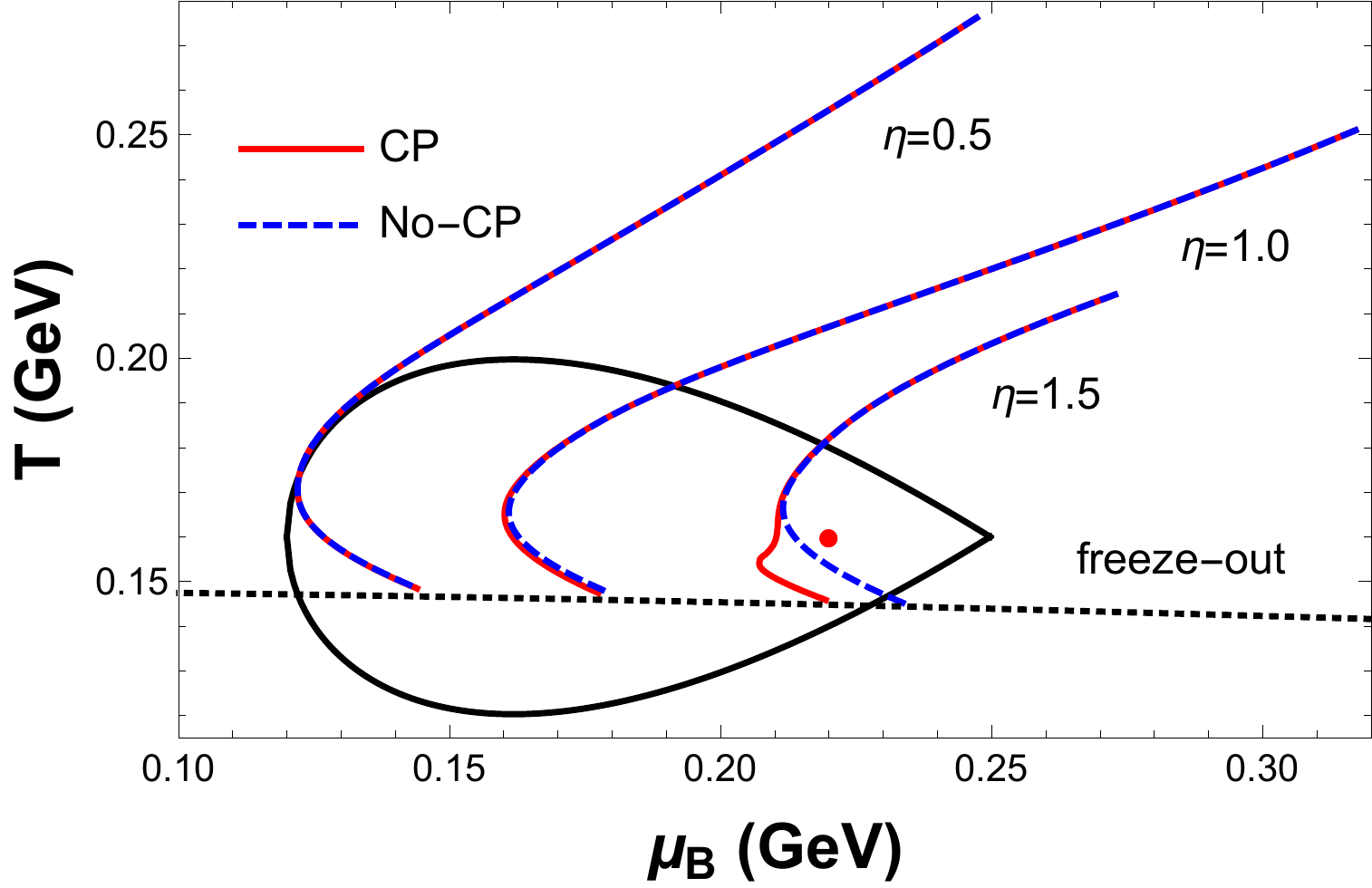}
\caption{(Color online) Hydrodynamic evolution trajectories in the $\mu_B$-$T$ plane for different
spatial rapidity, $\eta$, as well as with and without the presence of a critical
point.
Black solid curve plots the equal-$\xi$ contour, \emph{i.e.} the critical region, with the red dot illustrating the position of the critical point, see text. 
}
\label{fig:trj}
\end{figure}

To study the hydrodynamic evolution near the critical point we implement critical
behaviors of $\zeta$ and $\tau_{\sigma}$ within a chosen critical region. First, we
define the QCD critical region as the area enclosed by equal correlation length
contour $\xi(\mu_{B},T)=\xi_{0}$, where $\xi_{0}$ is the value of the correlation
length at the edge of the critical regime (c.f.~ Fig.~\ref{fig:trj}).  Within this critical region we use
\begin{equation}
\label{critical-scaling}
\zeta =\zeta_{0}\(\frac{\xi}{\xi_{0}}\)^3\, ,
\qquad \mathrm{and} \qquad
 \tau_{\Pi} =\tau^{0}_{\Pi}\(\frac{\xi}{\xi_{0}}\)^3 \, ,
\end{equation}
in accordance with Eq.~\eqref{zeta-all-relatioin}. $\zeta_0$ and $\tau_\Pi^0$ are the
bulk viscosity and bulk relaxation time outside the critical region, respectively.
Our choice for the $(T,\mu_B)$-dependence of $\zeta_0$ and $\tau_\Pi^0$ are
motivated by the holographic model based results of
Refs.~\cite{Buchel:2007mf,Natsuume:2007ty}
\bes
\begin{eqnarray}
\label{bulk-zero}
\zeta_0 &=& 
2 \bigg( \frac{1}{3}-c_s^2 \bigg) \frac{e+p}{4\pi T}\, , 
 \\
\label{tau-zero}
\tau_\Pi^0 &=& C_{\Pi} \frac{18-(9\ln 3 -\sqrt{3}\pi)}{24\pi T}\, .
\end{eqnarray}
\ees
To study the effects of relaxation time in more detail we consider hydrodynamic
evolutions with different choices of $C_{\Pi}$. 

Next, we model how $\xi$ varies over the $\mu_B$-$T$ plane, \emph{i.e.} $\xi(\mu_{B},
T)$. Near the Ising critical point the dependence of $\xi(r,h)$ on the Ising
variables, the reduced temperature $r$ and the re-scaled magnetic field $h$, is
universal. For convenience of the readers the details on $\xi(r,h)$ are provided
in Appendix.~\ref{sec:IS}. However, the mapping of the Ising variables $(r,h)$ onto
the thermodynamic variables $(\mu_B,T)$ is not universal.  For simplicity, we will
follow the widely-used prescription (see for
example~\cite{Berdnikov:1999ph,Nonaka:2004pg,Mukherjee:2015swa})
\be
\label{eq:Tmu_mapping}
\frac{T- T_c}{\Delta T}
=\frac{h}{\Delta h}\, ,
\qquad
\frac{\mu_{B}-\mu^{c}_{B}}{\Delta \mu_{B}}
= -\frac{r}{\Delta r}\, .
\ee
Here, $(\mu_B^c,T_c)$ is the location of the QCD critical point, and
$(\Delta\mu_B,\Delta T)$ is the width of the critical region in the $\mu_B$-$T$ plane.
The corresponding  width of the critical region in term of the Ising variables
$(\Delta r, \Delta h)$ is defined to be
\be
\xi(r=\Delta r, h=0)=\xi(r=0,h=\Delta h)
= \xi_{0}\, .
\ee
Specifically, the width of the QCD critical region is chosen to be  $(\Delta \mu_B,
\Delta T)= (0.1, 0.02)$~GeV, surrounding the critical point  located at
$(\mu^{c}_{B},T_{c} )= (0.22, 0.16 )$~GeV.  For numerical convenience, we set a upper
limit for $\xi$, \emph{i.e.} $\xi_{\rm max}=10\xi_{0}$. One might wonder that our
choice of $\mu^{c}_{B},T_{c}$ are unreasonable based upon our current knowledge of
the QCD phase diagram.  However, for our illustrative purpose, this choice is a
simple convenient way to make comparisons between evaluations with and without the
presence of a critical point, keeping other inputs (initial condition, EoS
\emph{etc.}) unchanged. The boundary of the critical region in the $\mu_{B}$-$T$
plane is illustrated in Fig.~\ref{fig:trj}. 

Finally, we end the hydrodynamic evolution and compute particle distributions using
the Cooper-Frye formalism on a (thermal) freeze-out surface characterized by a constant energy
density $\e_f = 0.25$ GeV/fm$^3$. This corresponds to freeze-out temperature $T_f
\sim 0.15$ GeV at vanishing chemical potential. The freeze-out curve in the
$\mu_{B}$-$T$ plane is also shown in Fig.~\ref{fig:trj}.

The bulk viscosity modifies the one-particle phase-space distribution in the Cooper-Frye formula. This distortion of the distribution $\delta f$ is determined using the Grad's moment expansion and the self-consistency conditions that off-equilibrium components of the energy-momentum tensors and net baryon number currents in kinetic theory and hydrodynamics match respectively \cite{Israel:1979wp}. See Ref.~\cite{Monnai:2009ad,*Monnai:2010qp} for the specific form.

\section{Results}
\label{sec:results}

\subsection{Trajectories in the $\mu_B$-$T$ plane}

The trajectories in the $\mu_B$-$T$ plane resulted from our hydrodynamic evolution,
with $\tau_\Pi$ corresponding to $C_{\Pi}=1$, are shown in Fig.~\ref{fig:trj}. The
trajectories for the CP and no-CP scenarios are nearly identical except for those
corresponding to $\eta=1.5$, \emph{i.e.} the one closest to the critical point. The
large value of $\zeta$ in the vicinity of the critical point produces more entropy,
and `pushes' the trajectory towards smaller values of $\mu_B$. Consequently,
trajectories passing the crossover side of the critical regime move further away from
the critical point.

\subsection{Bulk viscous pressure and flow}

\begin{figure}
\centering
\subfigure[]
{
\includegraphics[width=0.45\textwidth]{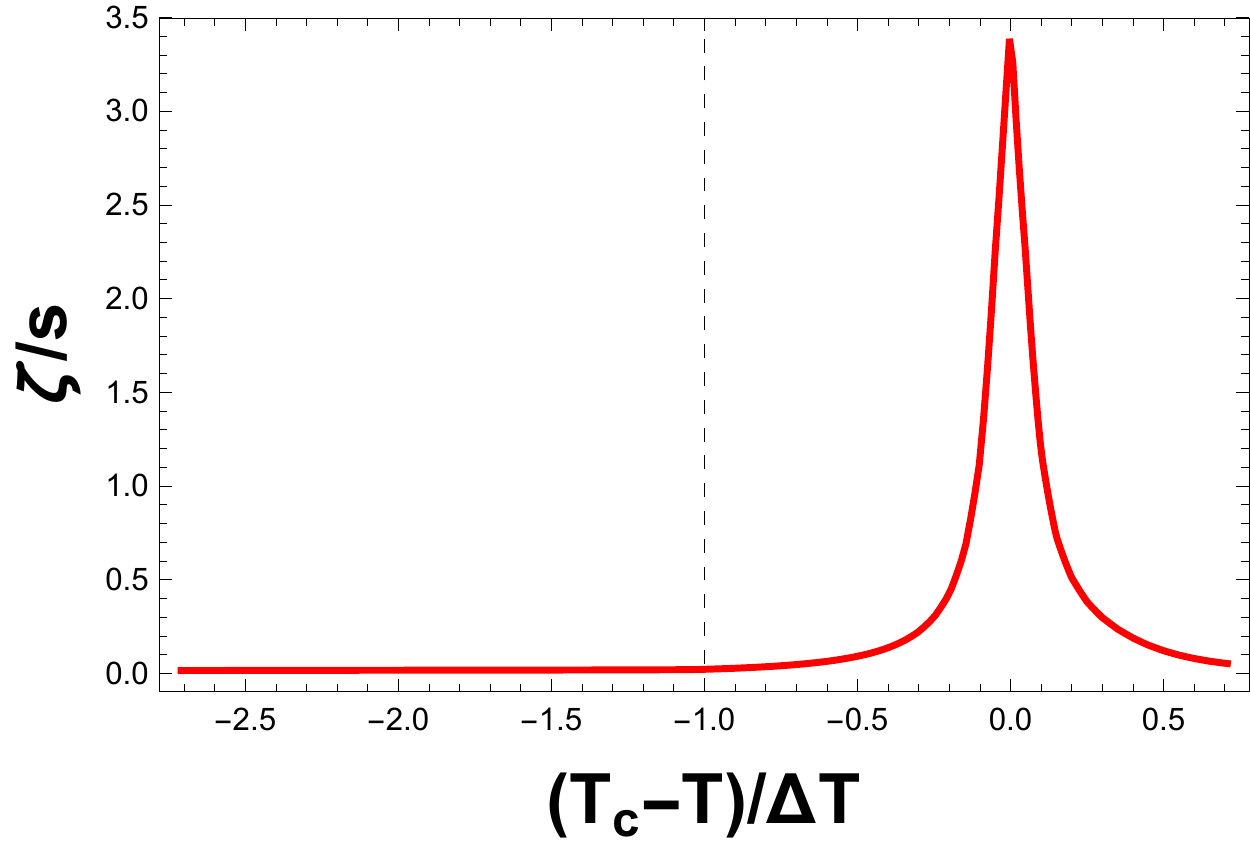}
\label{fig:zetaplot}
}
\subfigure[]
{
\includegraphics[width=0.45\textwidth]{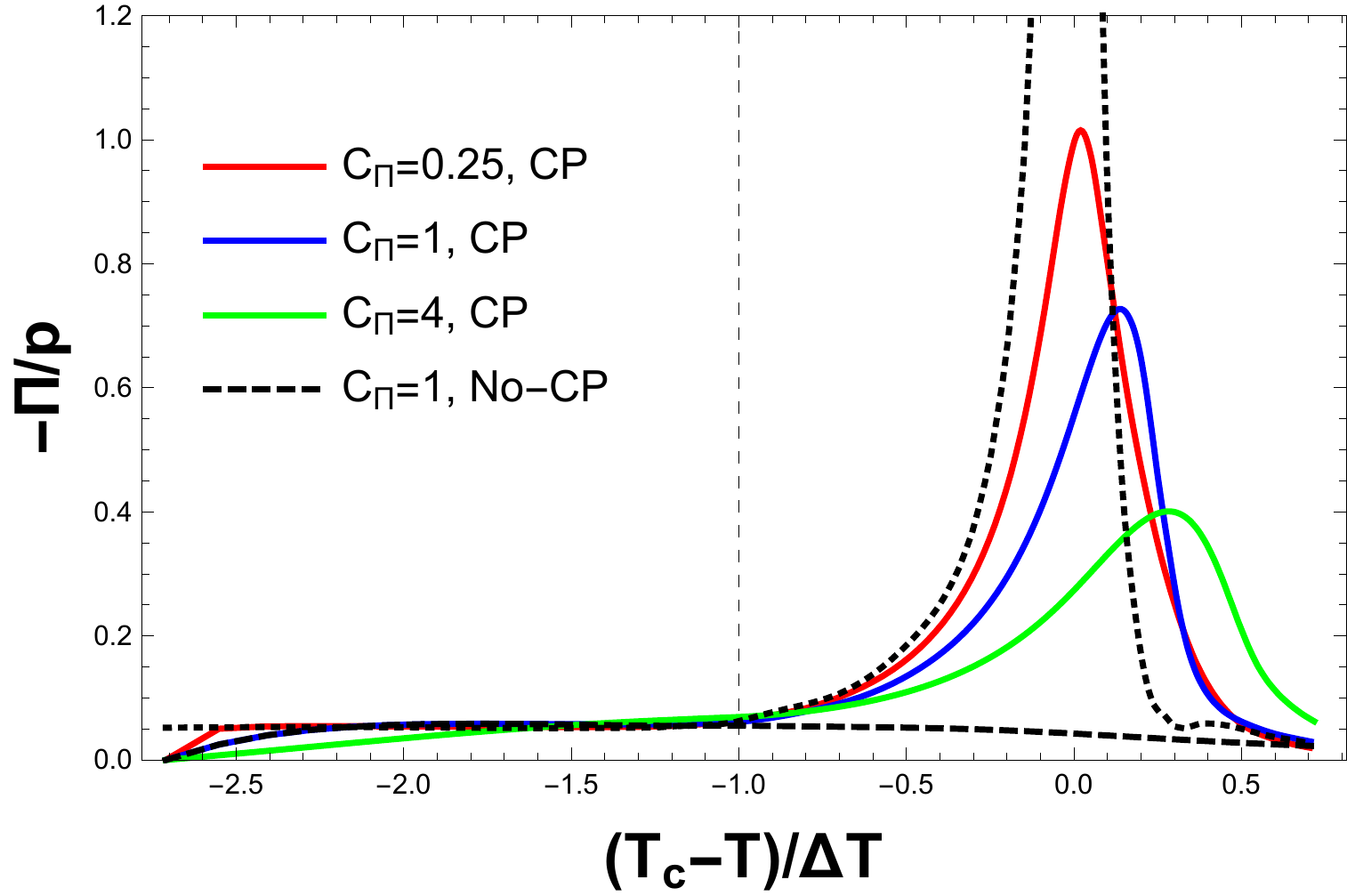}
\label{fig:Pi}
}
\subfigure[]
{
\includegraphics[width=.45\textwidth]{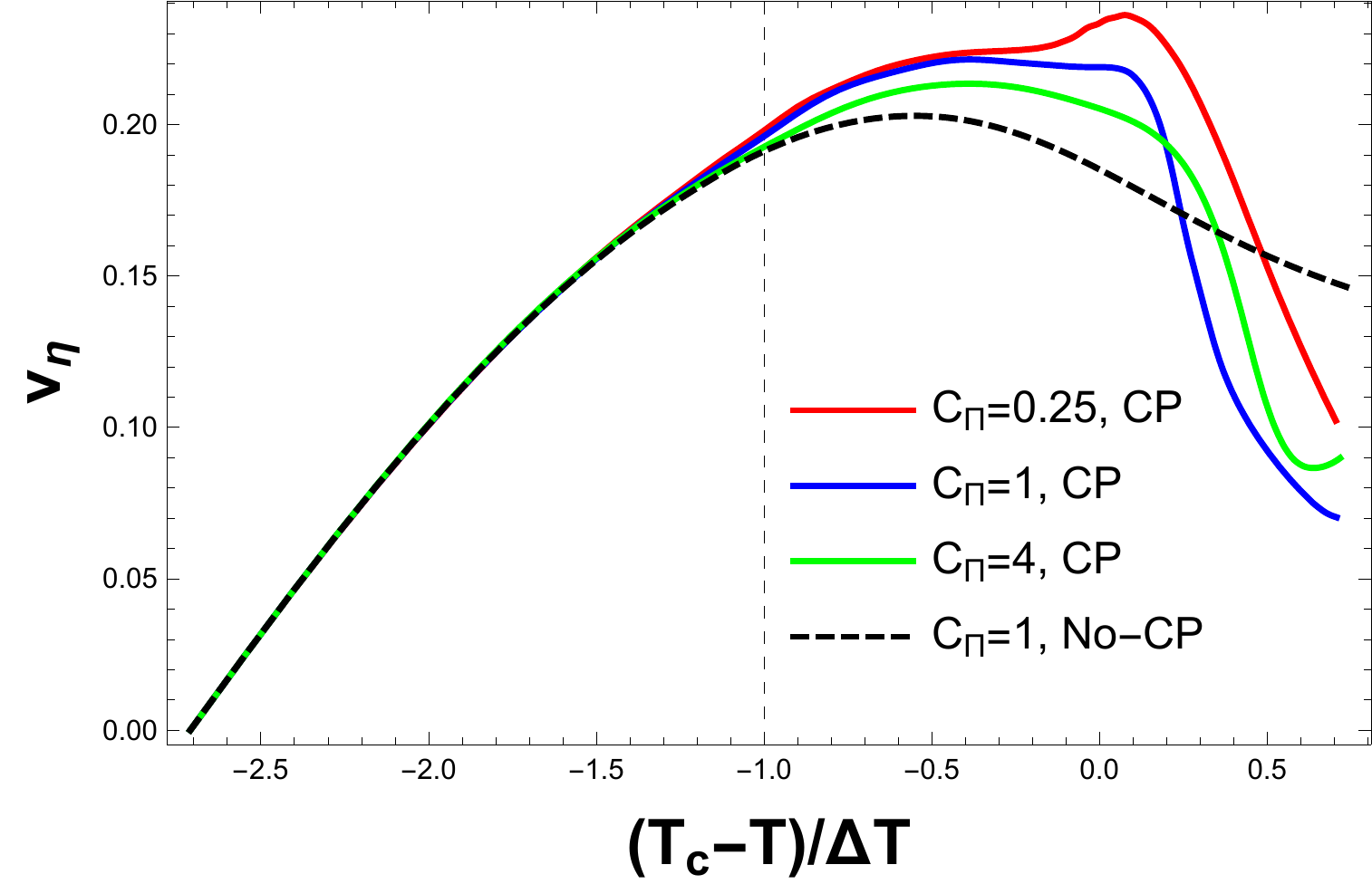}
\label{fig:flow}
}
\caption{(Color online) The ratios of bulk viscosity to entropy density $\zeta/s$ (a), bulk viscous
pressure to equilibrium pressure $\Pi/p$ (b), and the fluid velocity
$v_{\eta}=u^{\eta}/u^{\tau}$ (c) along the hydrodynamic evolution trajectory
corresponding to spatial rapidity $\eta=1.5$ and in presence of a critical point.
Results for evolution with several values of the bulk relaxation time $\tau_\Pi$
corresponding to $C_{\Pi}=0.25,1$ and $4$ (see Eq.~\ref{tau-zero}) are shown by red,
blue and green curves, respectively. The Navier-Stokes limit along the same
trajectory in the presence of the critical point also is shown by the black dotted
curve. The black dashed curves show results in absence of a critical point and with
$C_\pi=1$.
}
\label{fig:fixed-eta}
\end{figure}

We now turn to bulk viscous pressure by focusing on the evolution of $\Pi$ along the
trajectory corresponding to $\eta = 1.5$ for the CP scenario. First, in
Fig.~\ref{fig:zetaplot} we show the critical enhancement of the relevant
dimensionless quantity $\zeta/s$ along this trajectory, $s$ being the entropy
density. This enhanced bulk viscosity leads to a dramatic reduction of the bulk
viscous pressure, as shown in Fig.~\ref{fig:Pi}. For reference, in the same figure we
also show the corresponding Navier-Stokes value (black dotted line) and the no-CP
scenario (black dashed line). The growth of Naiver-Stokes value, $\zeta
\(\pd_{\mu}u^{\mu}\)$, in the critical regime is in accordance with the growth of
$\zeta$. While for no-CP $|\Pi|$ is an order of magnitude smaller than the
equilibrium pressure, for CP $|\Pi|$ can become comparable to the equilibrium
pressure within the critical regime. 

To clarify the role of $\tau_{\Pi}$ in the critical regime we solve Eq.~\eqref{IS-eq}
for several values of $\tau_\Pi$, corresponding to different choices of $C_\Pi$.
These results are shown in Fig.~\ref{fig:Pi}. As expected, a larger bulk relaxation
time not only reduces the maximal value $|\Pi|$ but also delays the growth. Note
that, such delayed growths make $|\Pi|$ substantially larger than its Navier-Stokes
limits at later times.  Similar observations regarding finite-time effects were made
in previous studies of on critical
fluctuations~\cite{Berdnikov:1999ph,Mukherjee:2015swa}; 
for off-equilibrium critical universal behavior induced by finite-time effects, see Ref.~\cite{Mukherjee:2016kyu}. 

It is easy to infer the effective pressure, $p_{\eff}=p+\Pi$, from the $\Pi/p$ in
Fig~\ref{fig:Pi}. While for no-CP $p_{\eff}$ is not much different from the equilibrium
pressure, in the presence of a critical point $p_{\eff}$ can be significantly suppressed
and become nearly vanishing depending on the largeness of $\tau_\Pi$.

In Fig.~\ref{fig:flow} we show the flow velocity, $v_{\eta}=u^{\eta}/u^{\tau}$ along
the same trajectory for different values of $\tau_\Pi$. Since the acceleration rate
of flow rapidity is proportional to the spatial gradients of $p_{\eff}$, for CP we observe
significantly large $v_\eta$ due the smaller $p_{\eff}$ at the forward-rapidity.

\subsection{Rapidity distributions of hadrons}

\begin{figure}[htb]
\center
\subfigure[]
{
\includegraphics[width=0.45\textwidth]{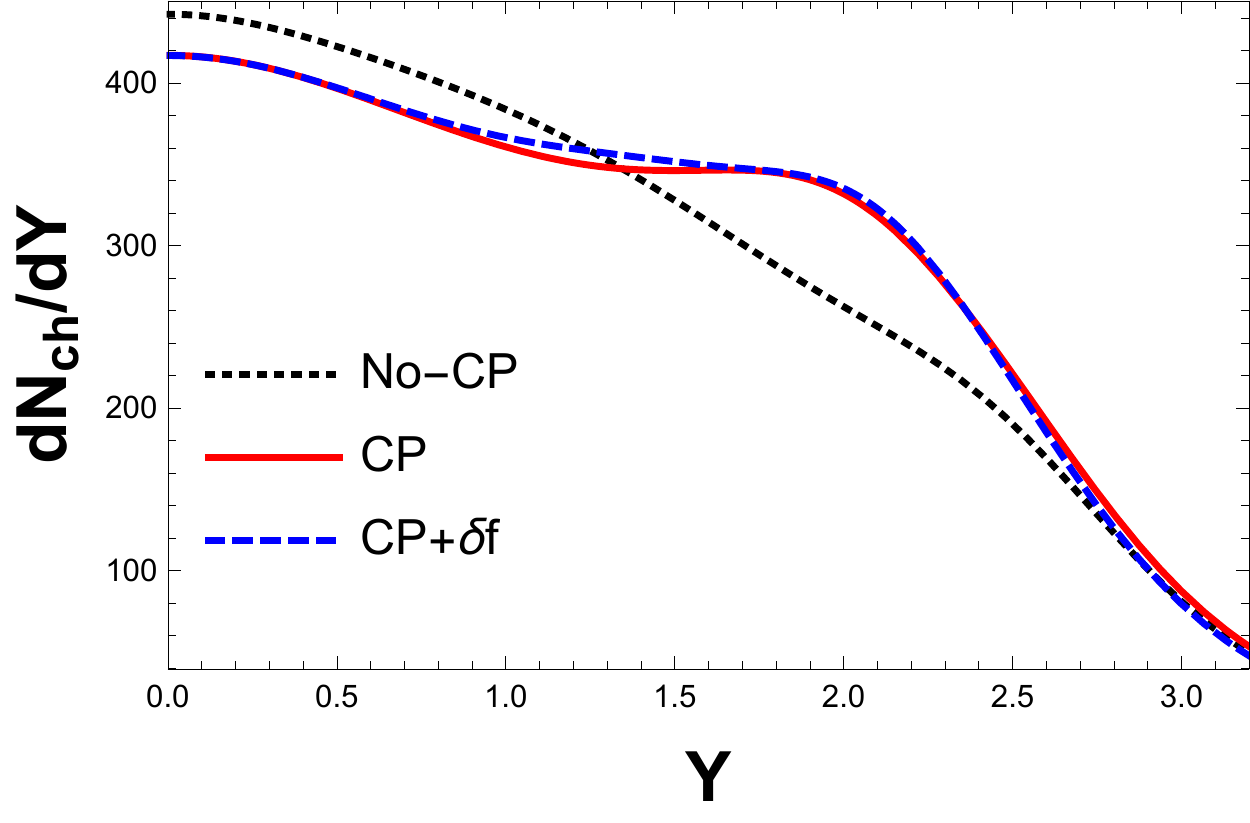}
\label{fig:Nch0}
}
\subfigure[]
{
\includegraphics[width=0.45\textwidth]{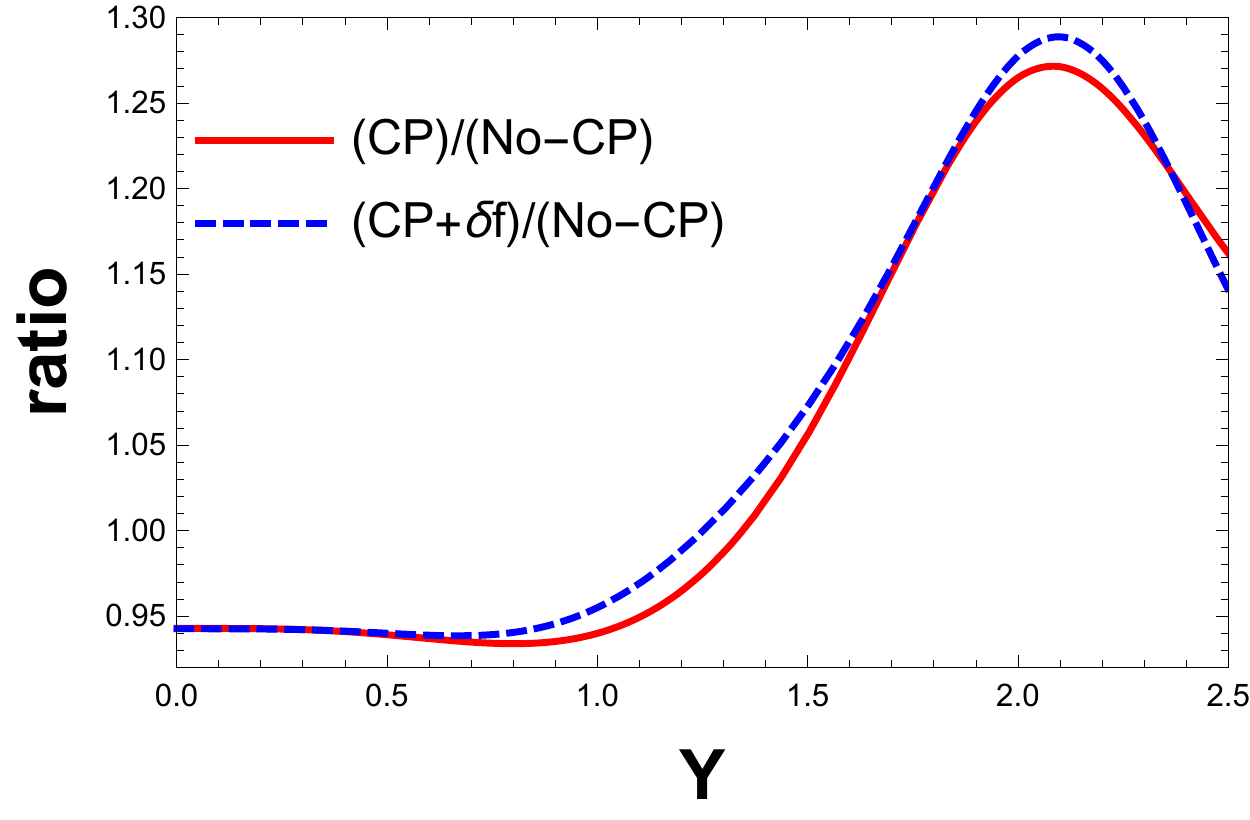}
\label{fig:Nchratio}
}
\caption{(Color online) (a) Charge particle multiplicity per unit momentum rapidity, $dN_{\rm
ch}/dY$, as a function of $Y$. The dashed black, solid red and dashed blue curves,
respectively, correspond to results  for hydrodynamical evolution with (CP) without
(no-CP) a critical point, as well as including the non-equilibrium $\delta f$
contribution to the freeze-out (CP+$\delta f$). (b) The relative change of $dN_{\rm
ch}/dY$ with and without a critical point. 
}
\label{fig:Nch}
\end{figure}

We now present couple observables accessible in the heavy-ion collision experiments.
We choose charged particle and net-baryon number multiplicities, $N_{\rm ch}$ and
$N_{\rm B}$ respectively, per unit momentum rapidity $Y$. The $dN_{\rm ch}/dY$ and
$dN_{\rm B}/dY$ are obtained at the freeze-out, carried out following the procedure
described before.  

Our results for $dN_{\rm ch}/dY$ are shown in Fig~\ref{fig:Nch}. In addition, we have
also found that the influence of the non-critical bulk viscosity, \emph{i.e.} in the
no-CP scenario, on $dN_{\rm ch}/dY$ without $\delta f$ is a mere few percent. For no-CP $dN_{\rm ch}/dY$
decreases monotonically with increasing $Y$, similar to the $\epsilon_{I}(\eta)$ in
Fig.~\ref{fig:EIC}.  In contrast, for the CP case $dN_{\rm ch}/dY$ becomes
non-monotonic in $Y$, showing a large increase around $Y=2$.  These charged particles
likely come from trajectories which are closer to the critical point, see
Fig.~\ref{fig:trj}. 
Here one should note that there is correlation between $\eta$ and $Y$ even after thermal smearing.
The enhanced particle production might be
understood intuitively by noting that growth of bulk viscosity \eqref{zeta_CEP} in
the vicinity of the QCD critical point induces an increase in entropy thus is 
accompanied by an increase in multiplicity.  The effects of enhanced entropy
production due to bulk viscosity near the critical point has also been discussed in
Ref.~\cite{Karsch:2007jc}. Additionally, the spatial gradient of $p_{\eff}$ is
steepened by the large bulk viscosity, accelerating the flow as indicated in
Fig.~\ref{fig:flow}. The entropy density is then carried from mid-rapidity to forward
rapidity because the entropy current is associated with the flow. This explains
slight reduction of multiplicity at mid-rapidity in Fig.~\ref{fig:Nch0}. As shown 
explicitly in Fig.~\ref{fig:Nchratio}, the presence of a critical point can enhance
$dN_{\rm ch}/dY$ in the forward-rapidity by as much as 30\%.

\begin{figure}[htb]
\center
\subfigure[]
{
\includegraphics[width=0.45\textwidth]{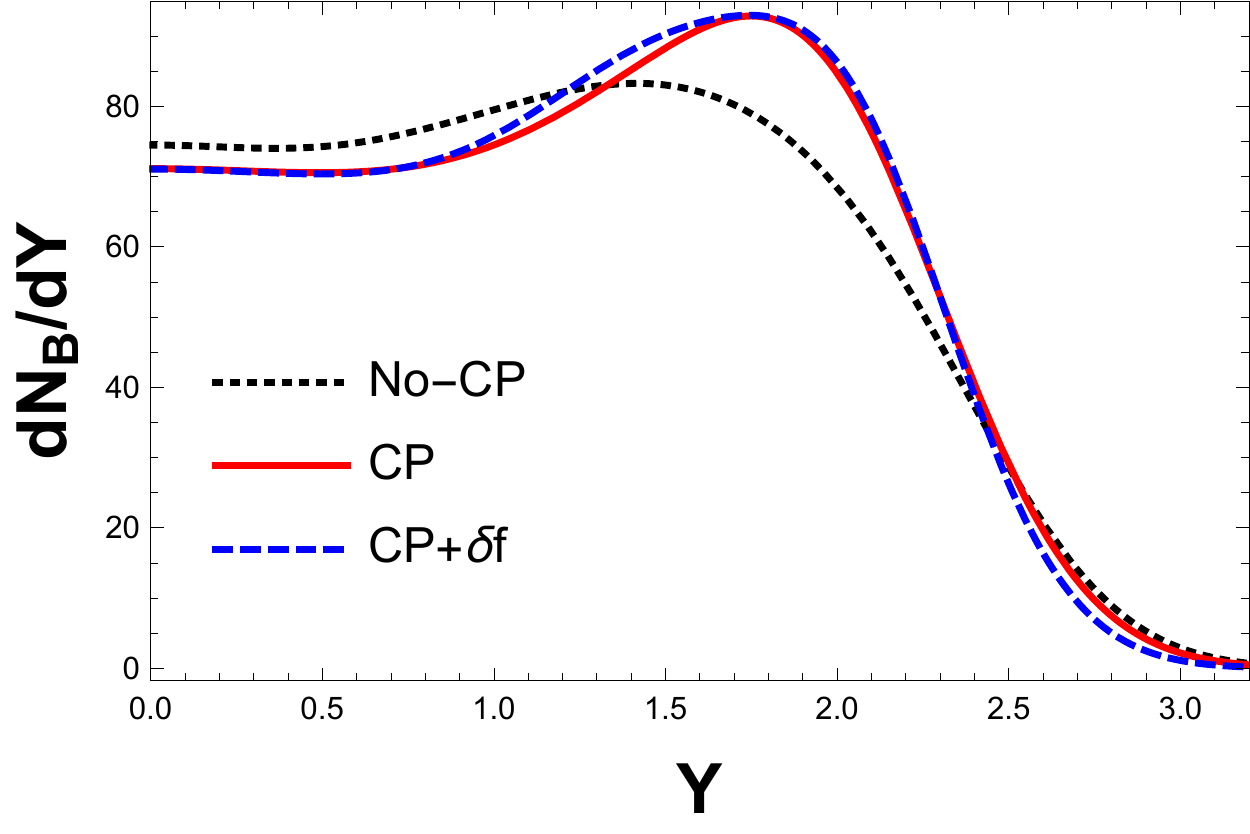}
\label{fig:dNBdy}
}
\subfigure[]
{
\includegraphics[width=0.45\textwidth]{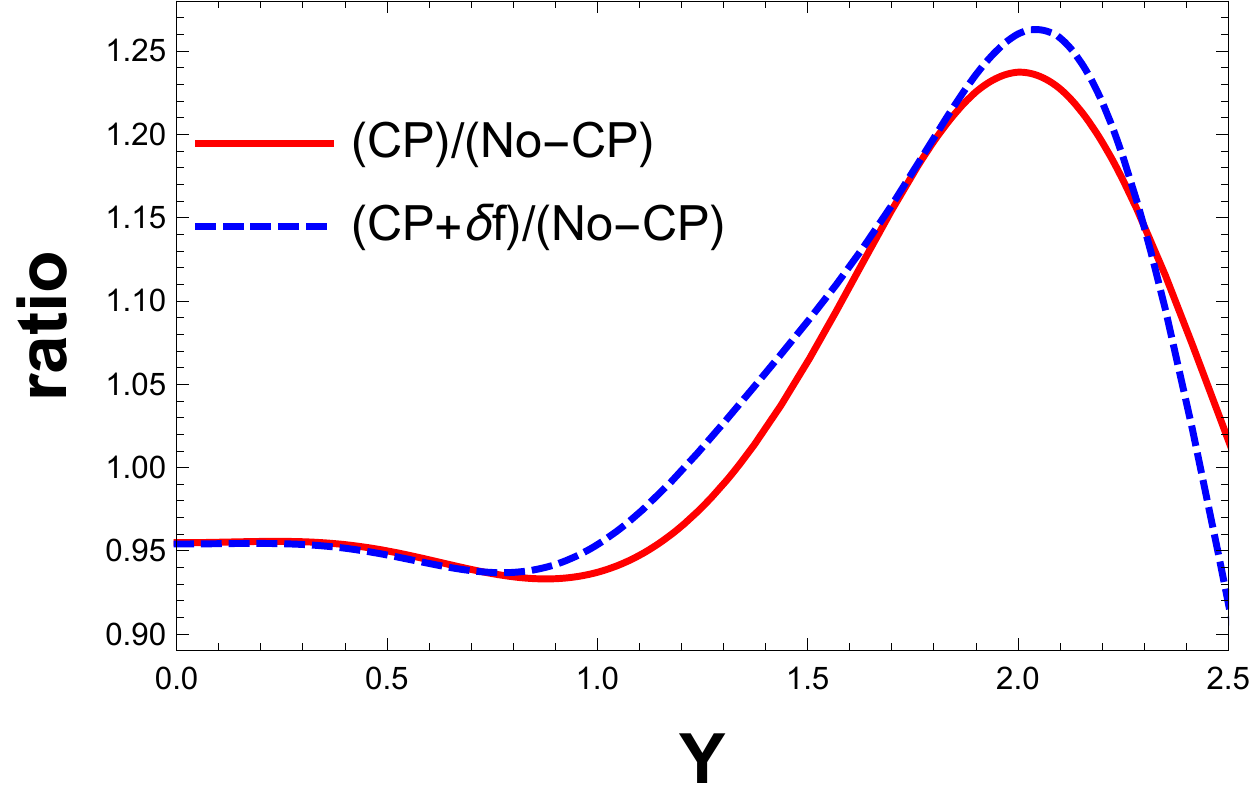}
\label{fig:NBratio}
}
\caption{(Color online) (a) Net-baryon multiplicity per unit momentum rapidity, $dN_{\rm B}/dY$, as
a function of $Y$. The dashed black, solid red and dashed blue curves, respectively,
correspond to results for hydrodynamical evolution with (CP) without (no-CP) a
critical point, as well as including the non-equilibrium $\delta f$ contribution to
the freeze-out (CP+$\delta f$). (b) The relative change of $dN_{\rm B}/dY$ with and
without a critical point.
}
\label{fig:NB}
\end{figure}

In Fig.~\ref{fig:NB} we present results for $dN_{\rm B}/dY$. The influence of a critical
point is similar to that in the case of charged particles. Since the entropy
production in presence of large viscosity will not change the net-baryon number, the
enhancement in $dN_{\rm B}/dY$ at the forward-rapidity is the effect of enhanced
convection caused by the larger spatial gradient in $p_{\eff}$. The relative
enhancement around $Y=2$ in Fig.~\ref{fig:NBratio} is, thus, smaller than that for
the $dN_{\rm ch}/dY$ in Fig.~\ref{fig:Nchratio}.

\section{ Summary and discussions }
\label{sec:summary}

This work is the first exploratory attempt to study the consequences of the
anticipated enhancement of bulk viscosity near the QCD critical point. We
incorporated the expected critical behaviors of the bulk viscosity and relaxation
time in Eq.~\eqref{zeta-all-relatioin} within a non-boost-invariant, longitudinally
expanding $1+1$ dimensional causal relativistic hydrodynamical evolution at non-zero
baryon density. To clarify the influence of the critically-enhanced bulk viscosity we
compared results from hydrodynamic evolution with and without the presence of a
critical point, but using the same initial conditions and the EoS.  As shown in
Figs.~\ref{fig:Nchratio}~and~\ref{fig:NBratio}, we found that the critically-enhanced
bulk viscosity led to a sizable increase of $dN_{\rm ch}/dY$ and $dN_{\rm B}/dY$ at
forward-rapidity; suggesting that particle spectra may contain discernible
information regarding the presence of a critical point in the QCD phase diagram.
Also, we have demonstrated that presence of a critical point will lead to a rapid
growth of the bulk viscous pressure, which in turn will soften the effective
pressure. In future, it will be interesting to explore whether such softening of the
effective pressure can lead to the observed non-monotonicity, as function of
$\sqrt{s}$, in the slope of the directed flow of net-protons \cite{Adamczyk:2014ipa}
or that in the re-scaled triangular flow \cite{Adamczyk:2016exq}, as these features
are sometime speculated to be related to the softening of the pressure. Of course,
such questions must be carefully addressed using state-of-the-art hydrodynamic
evolutions, and goes much beyond the scope of our present exploratory illustrative
study involving only longitudinal expansion, simplified initial conditions and
non-critical EoS. Use of state-of-the-art hydrodynamic evolution also will enable
access to transverse momentum spectra and azimuthal momentum anisotropy, which were
shown to be more sensitive to the bulk viscosity \cite{Song:2009rh,*Noronha-Hostler:2013gga, *Noronha-Hostler:2014dqa, *Ryu:2015vwa}.

We close with a short discussion on the applicability of the hydrodynamic with a
large bulk viscosity. Obviously, when $p_{\rm eff}<0$ hydrodynamical evolution
becomes mechanically unstable against cavitation, as has been discussed in a number
of references
\cite{Torrieri:2007fb,*Torrieri:2008ip,*Rajagopal:2009yw,*Habich:2014tpa} previously.
Indeed, if $\Pi$ is close to its corresponding Navier-Stokes value, the cavitation
seems unavoidable for trajectories near the critical point.  However, as we have
already discussed, the presence of $\tau_{\Pi}$ limits the growth of $\Pi$.
Consequently, in our current case, cavitation may only happen for very small values
of $\tau_\Pi$ corresponding to $C_\Pi<0.25$, see for example the solid red curve in
Fig.~\ref{fig:Pi}.  Furthermore, as pointed out in Ref.~\cite{Song:2009rh}, when the
viscosity is too large then one also needs to worry about the applicability of the
\IS formalism, namely the bulk viscous corrections to the energy-momentum tensor must
be much smaller compared to the ideal fluid terms, $|\Pi|/(\e+p) \ll 1$.  When
$|\Pi|\sim p$, one has $|\Pi|/(\e+p) \sim 1/(1+c^{-2}_{s})$. 
For $\mu_B=0$ and around the transition temperature $T_c$, the value of $c_s^2$ 
is about $0.15$ \cite{Bazavov:2014pvz}, giving $|\Pi|/(\epsilon+p) \sim 0.1$. Closer to the critical 
point and around the cross-over line the equation of state is expected to become even softer, 
leading to an even smaller value of $c_s^2$, and thus extending the the range of 
applicability of Israel-Stewart formalism.

One can also see
in Figs.~\ref{fig:Nch} and \ref{fig:NB} that the contributions of the bulk
viscous corrections $\delta f$ to the particle spectra are small even when $|\Pi|\sim
p$.

\acknowledgments
We would like to thank G.~Denicol, U.~Heinz, R.~Pisarski, L.~McLerran, K.~Rajagopal,
P.~Sorensen, and M.~Stephanov for very valuable discussions and M.~Nahrgang, T.~Sch\"{a}fer, and
B.~Schenke for commenting on the manuscript. AM was supported in part by RIKEN Special
Postdoctoral Researcher program and JSPS Postdoctoral Fellowship for Research Abroad.
This material is based upon work supported by the U.S. Department of Energy, Office
of Science, Office of Nuclear Physics, under Contract No. DE-SC0012704, and within
the framework of the Beam Energy Scan Theory (BEST) Topical Collaboration.

\bibliography{bulk}

\begin{thebibliography}{60}%
\makeatletter
\providecommand \@ifxundefined [1]{%
 \@ifx{#1\undefined}
}%
\providecommand \@ifnum [1]{%
 \ifnum #1\expandafter \@firstoftwo
 \else \expandafter \@secondoftwo
 \fi
}%
\providecommand \@ifx [1]{%
 \ifx #1\expandafter \@firstoftwo
 \else \expandafter \@secondoftwo
 \fi
}%
\providecommand \natexlab [1]{#1}%
\providecommand \enquote  [1]{``#1''}%
\providecommand \bibnamefont  [1]{#1}%
\providecommand \bibfnamefont [1]{#1}%
\providecommand \citenamefont [1]{#1}%
\providecommand \href@noop [0]{\@secondoftwo}%
\providecommand \href [0]{\begingroup \@sanitize@url \@href}%
\providecommand \@href[1]{\@@startlink{#1}\@@href}%
\providecommand \@@href[1]{\endgroup#1\@@endlink}%
\providecommand \@sanitize@url [0]{\catcode `\\12\catcode `\$12\catcode
  `\&12\catcode `\#12\catcode `\^12\catcode `\_12\catcode `\%12\relax}%
\providecommand \@@startlink[1]{}%
\providecommand \@@endlink[0]{}%
\providecommand \url  [0]{\begingroup\@sanitize@url \@url }%
\providecommand \@url [1]{\endgroup\@href {#1}{\urlprefix }}%
\providecommand \urlprefix  [0]{URL }%
\providecommand \Eprint [0]{\href }%
\providecommand \doibase [0]{http://dx.doi.org/}%
\providecommand \selectlanguage [0]{\@gobble}%
\providecommand \bibinfo  [0]{\@secondoftwo}%
\providecommand \bibfield  [0]{\@secondoftwo}%
\providecommand \translation [1]{[#1]}%
\providecommand \BibitemOpen [0]{}%
\providecommand \bibitemStop [0]{}%
\providecommand \bibitemNoStop [0]{.\EOS\space}%
\providecommand \EOS [0]{\spacefactor3000\relax}%
\providecommand \BibitemShut  [1]{\csname bibitem#1\endcsname}%
\let\auto@bib@innerbib\@empty
\bibitem [{\citenamefont {Stephanov}(2004)}]{Stephanov:2004wx}%
  \BibitemOpen
  \bibfield  {author} {\bibinfo {author} {\bibfnamefont {M.~A.}\ \bibnamefont
  {Stephanov}},\ }\href {\doibase 10.1142/S0217751X05027965} {\bibfield
  {journal} {\bibinfo  {journal} {Prog. Theor. Phys. Suppl.}\ }\textbf
  {\bibinfo {volume} {153}},\ \bibinfo {pages} {139} (\bibinfo {year}
  {2004})}\BibitemShut {NoStop}%
\bibitem [{\citenamefont {Stephanov}(2006)}]{Stephanov:2007fk}%
  \BibitemOpen
  \bibfield  {author} {\bibinfo {author} {\bibfnamefont {M.}~\bibnamefont
  {Stephanov}},\ }\href@noop {} {\bibfield  {journal} {\bibinfo  {journal}
  {PoS}\ }\textbf {\bibinfo {volume} {LAT2006}},\ \bibinfo {pages} {024}
  (\bibinfo {year} {2006})}\BibitemShut {NoStop}%
\bibitem [{\citenamefont {Fukushima}\ and\ \citenamefont
  {Hatsuda}(2011)}]{Fukushima:2010bq}%
  \BibitemOpen
  \bibfield  {author} {\bibinfo {author} {\bibfnamefont {K.}~\bibnamefont
  {Fukushima}}\ and\ \bibinfo {author} {\bibfnamefont {T.}~\bibnamefont
  {Hatsuda}},\ }\href {\doibase 10.1088/0034-4885/74/1/014001} {\bibfield
  {journal} {\bibinfo  {journal} {Rept. Prog. Phys.}\ }\textbf {\bibinfo
  {volume} {74}},\ \bibinfo {pages} {014001} (\bibinfo {year}
  {2011})}\BibitemShut {NoStop}%
\bibitem [{\citenamefont {Ding}\ \emph
  {et~al.}(2015{\natexlab{a}})\citenamefont {Ding}, \citenamefont {Karsch},\
  and\ \citenamefont {Mukherjee}}]{Ding:2015ona}%
  \BibitemOpen
  \bibfield  {author} {\bibinfo {author} {\bibfnamefont {H.-T.}\ \bibnamefont
  {Ding}}, \bibinfo {author} {\bibfnamefont {F.}~\bibnamefont {Karsch}}, \ and\
  \bibinfo {author} {\bibfnamefont {S.}~\bibnamefont {Mukherjee}},\ }\href
  {\doibase 10.1142/S0218301315300076} {\bibfield  {journal} {\bibinfo
  {journal} {Int. J. Mod. Phys.}\ }\textbf {\bibinfo {volume} {E24}},\ \bibinfo
  {pages} {1530007} (\bibinfo {year} {2015}{\natexlab{a}})}\BibitemShut
  {NoStop}%
\bibitem [{STA(2014)}]{STAR-wp}%
  \BibitemOpen
  \href@noop {} {\enquote {\bibinfo {title} {Studying the phase diagram of qcd
  matter at rhic},}\ } (\bibinfo {year} {2014}),\ \bibinfo {note}
  {\url{https://drupal.star.bnl.gov/STAR/files/BES_WPII_ver6.9_Cover.pdf}}\BibitemShut
  {NoStop}%
\bibitem [{\citenamefont {Heinz}\ \emph {et~al.}(2015)\citenamefont {Heinz},
  \citenamefont {Sorensen}, \citenamefont {Deshpande}, \citenamefont
  {Gagliardi}, \citenamefont {Karsch} \emph {et~al.}}]{Heinz:2015tua}%
  \BibitemOpen
  \bibfield  {author} {\bibinfo {author} {\bibfnamefont {U.}~\bibnamefont
  {Heinz}}, \bibinfo {author} {\bibfnamefont {P.}~\bibnamefont {Sorensen}},
  \bibinfo {author} {\bibfnamefont {A.}~\bibnamefont {Deshpande}}, \bibinfo
  {author} {\bibfnamefont {C.}~\bibnamefont {Gagliardi}}, \bibinfo {author}
  {\bibfnamefont {F.}~\bibnamefont {Karsch}},  \emph {et~al.},\ }\href@noop {}
  {\  (\bibinfo {year} {2015})},\ \Eprint {http://arxiv.org/abs/1501.06477}
  {arXiv:1501.06477 [nucl-th]} \BibitemShut {NoStop}%
\bibitem [{\citenamefont {Asakawa}\ and\ \citenamefont
  {Yazaki}(1989)}]{Asakawa:1989bq}%
  \BibitemOpen
  \bibfield  {author} {\bibinfo {author} {\bibfnamefont {M.}~\bibnamefont
  {Asakawa}}\ and\ \bibinfo {author} {\bibfnamefont {K.}~\bibnamefont
  {Yazaki}},\ }\href {\doibase 10.1016/0375-9474(89)90002-X} {\bibfield
  {journal} {\bibinfo  {journal} {Nucl. Phys.}\ }\textbf {\bibinfo {volume}
  {A504}},\ \bibinfo {pages} {668} (\bibinfo {year} {1989})}\BibitemShut
  {NoStop}%
\bibitem [{\citenamefont {Barducci}\ \emph {et~al.}(1989)\citenamefont
  {Barducci}, \citenamefont {Casalbuoni}, \citenamefont {De~Curtis},
  \citenamefont {Gatto},\ and\ \citenamefont {Pettini}}]{Barducci:1989wi}%
  \BibitemOpen
  \bibfield  {author} {\bibinfo {author} {\bibfnamefont {A.}~\bibnamefont
  {Barducci}}, \bibinfo {author} {\bibfnamefont {R.}~\bibnamefont
  {Casalbuoni}}, \bibinfo {author} {\bibfnamefont {S.}~\bibnamefont
  {De~Curtis}}, \bibinfo {author} {\bibfnamefont {R.}~\bibnamefont {Gatto}}, \
  and\ \bibinfo {author} {\bibfnamefont {G.}~\bibnamefont {Pettini}},\ }\href
  {\doibase 10.1016/0370-2693(89)90695-3} {\bibfield  {journal} {\bibinfo
  {journal} {Phys. Lett.}\ }\textbf {\bibinfo {volume} {B231}},\ \bibinfo
  {pages} {463} (\bibinfo {year} {1989})}\BibitemShut {NoStop}%
\bibitem [{\citenamefont {Halasz}\ \emph {et~al.}(1998)\citenamefont {Halasz},
  \citenamefont {Jackson}, \citenamefont {Shrock}, \citenamefont {Stephanov},\
  and\ \citenamefont {Verbaarschot}}]{Halasz:1998qr}%
  \BibitemOpen
  \bibfield  {author} {\bibinfo {author} {\bibfnamefont {A.~M.}\ \bibnamefont
  {Halasz}}, \bibinfo {author} {\bibfnamefont {A.}~\bibnamefont {Jackson}},
  \bibinfo {author} {\bibfnamefont {R.}~\bibnamefont {Shrock}}, \bibinfo
  {author} {\bibfnamefont {M.~A.}\ \bibnamefont {Stephanov}}, \ and\ \bibinfo
  {author} {\bibfnamefont {J.}~\bibnamefont {Verbaarschot}},\ }\href {\doibase
  10.1103/PhysRevD.58.096007} {\bibfield  {journal} {\bibinfo  {journal} {Phys.
  Rev.}\ }\textbf {\bibinfo {volume} {D58}},\ \bibinfo {pages} {096007}
  (\bibinfo {year} {1998})}\BibitemShut {NoStop}%
\bibitem [{\citenamefont {Berges}\ and\ \citenamefont
  {Rajagopal}(1999)}]{Berges:1998rc}%
  \BibitemOpen
  \bibfield  {author} {\bibinfo {author} {\bibfnamefont {J.}~\bibnamefont
  {Berges}}\ and\ \bibinfo {author} {\bibfnamefont {K.}~\bibnamefont
  {Rajagopal}},\ }\href {\doibase 10.1016/S0550-3213(98)00620-8} {\bibfield
  {journal} {\bibinfo  {journal} {Nucl. Phys.}\ }\textbf {\bibinfo {volume}
  {B538}},\ \bibinfo {pages} {215} (\bibinfo {year} {1999})}\BibitemShut
  {NoStop}%
\bibitem [{\citenamefont {Stephanov}\ \emph {et~al.}(1998)\citenamefont
  {Stephanov}, \citenamefont {Rajagopal},\ and\ \citenamefont
  {Shuryak}}]{Stephanov:1998dy}%
  \BibitemOpen
  \bibfield  {author} {\bibinfo {author} {\bibfnamefont {M.~A.}\ \bibnamefont
  {Stephanov}}, \bibinfo {author} {\bibfnamefont {K.}~\bibnamefont
  {Rajagopal}}, \ and\ \bibinfo {author} {\bibfnamefont {E.~V.}\ \bibnamefont
  {Shuryak}},\ }\href {\doibase 10.1103/PhysRevLett.81.4816} {\bibfield
  {journal} {\bibinfo  {journal} {Phys. Rev. Lett.}\ }\textbf {\bibinfo
  {volume} {81}},\ \bibinfo {pages} {4816} (\bibinfo {year}
  {1998})}\BibitemShut {NoStop}%
\bibitem [{\citenamefont {Stephanov}(2009)}]{Stephanov:2008qz}%
  \BibitemOpen
  \bibfield  {author} {\bibinfo {author} {\bibfnamefont {M.}~\bibnamefont
  {Stephanov}},\ }\href {\doibase 10.1103/PhysRevLett.102.032301} {\bibfield
  {journal} {\bibinfo  {journal} {Phys. Rev. Lett.}\ }\textbf {\bibinfo
  {volume} {102}},\ \bibinfo {pages} {032301} (\bibinfo {year}
  {2009})}\BibitemShut {NoStop}%
\bibitem [{\citenamefont {Hatta}\ and\ \citenamefont
  {Stephanov}(2003)}]{Hatta:2003wn}%
  \BibitemOpen
  \bibfield  {author} {\bibinfo {author} {\bibfnamefont {Y.}~\bibnamefont
  {Hatta}}\ and\ \bibinfo {author} {\bibfnamefont {M.}~\bibnamefont
  {Stephanov}},\ }\href {\doibase 10.1103/PhysRevLett.91.102003} {\bibfield
  {journal} {\bibinfo  {journal} {Phys. Rev. Lett.}\ }\textbf {\bibinfo
  {volume} {91}},\ \bibinfo {pages} {102003} (\bibinfo {year}
  {2003})}\BibitemShut {NoStop}%
\bibitem [{\citenamefont {Athanasiou}\ \emph {et~al.}(2010)\citenamefont
  {Athanasiou}, \citenamefont {Rajagopal},\ and\ \citenamefont
  {Stephanov}}]{Athanasiou:2010kw}%
  \BibitemOpen
  \bibfield  {author} {\bibinfo {author} {\bibfnamefont {C.}~\bibnamefont
  {Athanasiou}}, \bibinfo {author} {\bibfnamefont {K.}~\bibnamefont
  {Rajagopal}}, \ and\ \bibinfo {author} {\bibfnamefont {M.}~\bibnamefont
  {Stephanov}},\ }\href {\doibase 10.1103/PhysRevD.82.074008} {\bibfield
  {journal} {\bibinfo  {journal} {Phys. Rev.}\ }\textbf {\bibinfo {volume}
  {D82}},\ \bibinfo {pages} {074008} (\bibinfo {year} {2010})}\BibitemShut
  {NoStop}%
\bibitem [{\citenamefont {Son}\ and\ \citenamefont
  {Stephanov}(2004)}]{Son:2004iv}%
  \BibitemOpen
  \bibfield  {author} {\bibinfo {author} {\bibfnamefont {D.}~\bibnamefont
  {Son}}\ and\ \bibinfo {author} {\bibfnamefont {M.}~\bibnamefont
  {Stephanov}},\ }\href {\doibase 10.1103/PhysRevD.70.056001} {\bibfield
  {journal} {\bibinfo  {journal} {Phys. Rev.}\ }\textbf {\bibinfo {volume}
  {D70}},\ \bibinfo {pages} {056001} (\bibinfo {year} {2004})}\BibitemShut
  {NoStop}%
\bibitem [{\citenamefont {Fujii}(2003)}]{Fujii:2003bz}%
  \BibitemOpen
  \bibfield  {author} {\bibinfo {author} {\bibfnamefont {H.}~\bibnamefont
  {Fujii}},\ }\href {\doibase 10.1103/PhysRevD.67.094018} {\bibfield  {journal}
  {\bibinfo  {journal} {Phys. Rev.}\ }\textbf {\bibinfo {volume} {D67}},\
  \bibinfo {pages} {094018} (\bibinfo {year} {2003})}\BibitemShut {NoStop}%
\bibitem [{\citenamefont {Fujii}\ and\ \citenamefont
  {Ohtani}(2004{\natexlab{a}})}]{Fujii:2004za}%
  \BibitemOpen
  \bibfield  {author} {\bibinfo {author} {\bibfnamefont {H.}~\bibnamefont
  {Fujii}}\ and\ \bibinfo {author} {\bibfnamefont {M.}~\bibnamefont {Ohtani}},\
  }\href {\doibase 10.1143/PTPS.153.157} {\bibfield  {journal} {\bibinfo
  {journal} {Prog. Theor. Phys. Suppl.}\ }\textbf {\bibinfo {volume} {153}},\
  \bibinfo {pages} {157} (\bibinfo {year} {2004}{\natexlab{a}})}\BibitemShut
  {NoStop}%
\bibitem [{\citenamefont {Fujii}\ and\ \citenamefont
  {Ohtani}(2004{\natexlab{b}})}]{Fujii:2004jt}%
  \BibitemOpen
  \bibfield  {author} {\bibinfo {author} {\bibfnamefont {H.}~\bibnamefont
  {Fujii}}\ and\ \bibinfo {author} {\bibfnamefont {M.}~\bibnamefont {Ohtani}},\
  }\href {\doibase 10.1103/PhysRevD.70.014016} {\bibfield  {journal} {\bibinfo
  {journal} {Phys. Rev.}\ }\textbf {\bibinfo {volume} {D70}},\ \bibinfo {pages}
  {014016} (\bibinfo {year} {2004}{\natexlab{b}})}\BibitemShut {NoStop}%
\bibitem [{\citenamefont {Hohenberg}\ and\ \citenamefont
  {Halperin}(1977)}]{RevModPhys.49.435}%
  \BibitemOpen
  \bibfield  {author} {\bibinfo {author} {\bibfnamefont {P.~C.}\ \bibnamefont
  {Hohenberg}}\ and\ \bibinfo {author} {\bibfnamefont {B.~I.}\ \bibnamefont
  {Halperin}},\ }\href {\doibase 10.1103/RevModPhys.49.435} {\bibfield
  {journal} {\bibinfo  {journal} {Rev. Mod. Phys.}\ }\textbf {\bibinfo {volume}
  {49}},\ \bibinfo {pages} {435} (\bibinfo {year} {1977})}\BibitemShut
  {NoStop}%
\bibitem [{\citenamefont {Moore}\ and\ \citenamefont
  {Saremi}(2008)}]{Moore:2008ws}%
  \BibitemOpen
  \bibfield  {author} {\bibinfo {author} {\bibfnamefont {G.~D.}\ \bibnamefont
  {Moore}}\ and\ \bibinfo {author} {\bibfnamefont {O.}~\bibnamefont {Saremi}},\
  }\href {\doibase 10.1088/1126-6708/2008/09/015} {\bibfield  {journal}
  {\bibinfo  {journal} {JHEP}\ }\textbf {\bibinfo {volume} {09}},\ \bibinfo
  {pages} {015} (\bibinfo {year} {2008})}\BibitemShut {NoStop}%
\bibitem [{\citenamefont {Onuki}(1997)}]{PhysRevE.55.403}%
  \BibitemOpen
  \bibfield  {author} {\bibinfo {author} {\bibfnamefont {A.}~\bibnamefont
  {Onuki}},\ }\href {\doibase 10.1103/PhysRevE.55.403} {\bibfield  {journal}
  {\bibinfo  {journal} {Phys. Rev. E}\ }\textbf {\bibinfo {volume} {55}},\
  \bibinfo {pages} {403} (\bibinfo {year} {1997})}\BibitemShut {NoStop}%
\bibitem [{\citenamefont {Paech}\ \emph {et~al.}(2003)\citenamefont {Paech},
  \citenamefont {Stoecker},\ and\ \citenamefont {Dumitru}}]{Paech:2003fe}%
  \BibitemOpen
  \bibfield  {author} {\bibinfo {author} {\bibfnamefont {K.}~\bibnamefont
  {Paech}}, \bibinfo {author} {\bibfnamefont {H.}~\bibnamefont {Stoecker}}, \
  and\ \bibinfo {author} {\bibfnamefont {A.}~\bibnamefont {Dumitru}},\ }\href
  {\doibase 10.1103/PhysRevC.68.044907} {\bibfield  {journal} {\bibinfo
  {journal} {Phys. Rev.}\ }\textbf {\bibinfo {volume} {C68}},\ \bibinfo {pages}
  {044907} (\bibinfo {year} {2003})}\BibitemShut {NoStop}%
\bibitem [{\citenamefont {Nonaka}\ and\ \citenamefont
  {Asakawa}(2005)}]{Nonaka:2004pg}%
  \BibitemOpen
  \bibfield  {author} {\bibinfo {author} {\bibfnamefont {C.}~\bibnamefont
  {Nonaka}}\ and\ \bibinfo {author} {\bibfnamefont {M.}~\bibnamefont
  {Asakawa}},\ }\href {\doibase 10.1103/PhysRevC.71.044904} {\bibfield
  {journal} {\bibinfo  {journal} {Phys. Rev.}\ }\textbf {\bibinfo {volume}
  {C71}},\ \bibinfo {pages} {044904} (\bibinfo {year} {2005})}\BibitemShut
  {NoStop}%
\bibitem [{\citenamefont {Paech}\ and\ \citenamefont
  {Dumitru}(2005)}]{Paech:2005cx}%
  \BibitemOpen
  \bibfield  {author} {\bibinfo {author} {\bibfnamefont {K.}~\bibnamefont
  {Paech}}\ and\ \bibinfo {author} {\bibfnamefont {A.}~\bibnamefont
  {Dumitru}},\ }\href {\doibase 10.1016/j.physletb.2005.08.006} {\bibfield
  {journal} {\bibinfo  {journal} {Phys. Lett.}\ }\textbf {\bibinfo {volume}
  {B623}},\ \bibinfo {pages} {200} (\bibinfo {year} {2005})}\BibitemShut
  {NoStop}%
\bibitem [{\citenamefont {Herold}\ \emph {et~al.}(2013)\citenamefont {Herold},
  \citenamefont {Nahrgang}, \citenamefont {Mishustin},\ and\ \citenamefont
  {Bleicher}}]{Herold:2013bi}%
  \BibitemOpen
  \bibfield  {author} {\bibinfo {author} {\bibfnamefont {C.}~\bibnamefont
  {Herold}}, \bibinfo {author} {\bibfnamefont {M.}~\bibnamefont {Nahrgang}},
  \bibinfo {author} {\bibfnamefont {I.}~\bibnamefont {Mishustin}}, \ and\
  \bibinfo {author} {\bibfnamefont {M.}~\bibnamefont {Bleicher}},\ }\href
  {\doibase 10.1103/PhysRevC.87.014907} {\bibfield  {journal} {\bibinfo
  {journal} {Phys. Rev.}\ }\textbf {\bibinfo {volume} {C87}},\ \bibinfo {pages}
  {014907} (\bibinfo {year} {2013})}\BibitemShut {NoStop}%
\bibitem [{\citenamefont {Nahrgang}(2016)}]{Nahrgang:2016ayr}%
  \BibitemOpen
  \bibfield  {author} {\bibinfo {author} {\bibfnamefont {M.}~\bibnamefont
  {Nahrgang}},\ }\href@noop {} {\  (\bibinfo {year} {2016})},\ \Eprint
  {http://arxiv.org/abs/1601.07437} {arXiv:1601.07437 [nucl-th]} \BibitemShut
  {NoStop}%
\bibitem [{\citenamefont {Kapusta}\ and\ \citenamefont
  {Torres-Rincon}(2012)}]{Kapusta:2012zb}%
  \BibitemOpen
  \bibfield  {author} {\bibinfo {author} {\bibfnamefont {J.~I.}\ \bibnamefont
  {Kapusta}}\ and\ \bibinfo {author} {\bibfnamefont {J.~M.}\ \bibnamefont
  {Torres-Rincon}},\ }\href {\doibase 10.1103/PhysRevC.86.054911} {\bibfield
  {journal} {\bibinfo  {journal} {Phys. Rev.}\ }\textbf {\bibinfo {volume}
  {C86}},\ \bibinfo {pages} {054911} (\bibinfo {year} {2012})}\BibitemShut
  {NoStop}%
\bibitem [{\citenamefont {Israel}\ and\ \citenamefont
  {Stewart}(1979)}]{Israel:1979wp}%
  \BibitemOpen
  \bibfield  {author} {\bibinfo {author} {\bibfnamefont {W.}~\bibnamefont
  {Israel}}\ and\ \bibinfo {author} {\bibfnamefont {J.~M.}\ \bibnamefont
  {Stewart}},\ }\href {\doibase 10.1016/0003-4916(79)90130-1} {\bibfield
  {journal} {\bibinfo  {journal} {Annals Phys.}\ }\textbf {\bibinfo {volume}
  {118}},\ \bibinfo {pages} {341} (\bibinfo {year} {1979})}\BibitemShut
  {NoStop}%
\bibitem [{\citenamefont {Romatschke}(2010)}]{Romatschke:2009im}%
  \BibitemOpen
  \bibfield  {author} {\bibinfo {author} {\bibfnamefont {P.}~\bibnamefont
  {Romatschke}},\ }\href {\doibase 10.1142/S0218301310014613} {\bibfield
  {journal} {\bibinfo  {journal} {Int. J. Mod. Phys.}\ }\textbf {\bibinfo
  {volume} {E19}},\ \bibinfo {pages} {1} (\bibinfo {year} {2010})}\BibitemShut
  {NoStop}%
\bibitem [{\citenamefont {Monnai}(2012)}]{Monnai:2012jc}%
  \BibitemOpen
  \bibfield  {author} {\bibinfo {author} {\bibfnamefont {A.}~\bibnamefont
  {Monnai}},\ }\href {\doibase 10.1103/PhysRevC.86.014908} {\bibfield
  {journal} {\bibinfo  {journal} {Phys. Rev.}\ }\textbf {\bibinfo {volume}
  {C86}},\ \bibinfo {pages} {014908} (\bibinfo {year} {2012})}\BibitemShut
  {NoStop}%
\bibitem [{\citenamefont {Drescher}\ and\ \citenamefont
  {Nara}(2007{\natexlab{a}})}]{Drescher:2006ca}%
  \BibitemOpen
  \bibfield  {author} {\bibinfo {author} {\bibfnamefont {H.~J.}\ \bibnamefont
  {Drescher}}\ and\ \bibinfo {author} {\bibfnamefont {Y.}~\bibnamefont
  {Nara}},\ }\href {\doibase 10.1103/PhysRevC.75.034905} {\bibfield  {journal}
  {\bibinfo  {journal} {Phys. Rev.}\ }\textbf {\bibinfo {volume} {C75}},\
  \bibinfo {pages} {034905} (\bibinfo {year} {2007}{\natexlab{a}})}\BibitemShut
  {NoStop}%
\bibitem [{\citenamefont {Drescher}\ and\ \citenamefont
  {Nara}(2007{\natexlab{b}})}]{Drescher:2007ax}%
  \BibitemOpen
  \bibfield  {author} {\bibinfo {author} {\bibfnamefont {H.-J.}\ \bibnamefont
  {Drescher}}\ and\ \bibinfo {author} {\bibfnamefont {Y.}~\bibnamefont
  {Nara}},\ }\href {\doibase 10.1103/PhysRevC.76.041903} {\bibfield  {journal}
  {\bibinfo  {journal} {Phys. Rev.}\ }\textbf {\bibinfo {volume} {C76}},\
  \bibinfo {pages} {041903} (\bibinfo {year} {2007}{\natexlab{b}})}\BibitemShut
  {NoStop}%
\bibitem [{\citenamefont {Mehtar-Tani}\ and\ \citenamefont
  {Wolschin}(2009{\natexlab{a}})}]{MehtarTani:2008qg}%
  \BibitemOpen
  \bibfield  {author} {\bibinfo {author} {\bibfnamefont {Y.}~\bibnamefont
  {Mehtar-Tani}}\ and\ \bibinfo {author} {\bibfnamefont {G.}~\bibnamefont
  {Wolschin}},\ }\href {\doibase 10.1103/PhysRevLett.102.182301} {\bibfield
  {journal} {\bibinfo  {journal} {Phys. Rev. Lett.}\ }\textbf {\bibinfo
  {volume} {102}},\ \bibinfo {pages} {182301} (\bibinfo {year}
  {2009}{\natexlab{a}})}\BibitemShut {NoStop}%
\bibitem [{\citenamefont {Mehtar-Tani}\ and\ \citenamefont
  {Wolschin}(2009{\natexlab{b}})}]{MehtarTani:2009dv}%
  \BibitemOpen
  \bibfield  {author} {\bibinfo {author} {\bibfnamefont {Y.}~\bibnamefont
  {Mehtar-Tani}}\ and\ \bibinfo {author} {\bibfnamefont {G.}~\bibnamefont
  {Wolschin}},\ }\href {\doibase 10.1103/PhysRevC.80.054905} {\bibfield
  {journal} {\bibinfo  {journal} {Phys. Rev.}\ }\textbf {\bibinfo {volume}
  {C80}},\ \bibinfo {pages} {054905} (\bibinfo {year}
  {2009}{\natexlab{b}})}\BibitemShut {NoStop}%
\bibitem [{\citenamefont {Bazavov}\ \emph {et~al.}(2014)\citenamefont {Bazavov}
  \emph {et~al.}}]{Bazavov:2014pvz}%
  \BibitemOpen
  \bibfield  {author} {\bibinfo {author} {\bibfnamefont {A.}~\bibnamefont
  {Bazavov}} \emph {et~al.} (\bibinfo {collaboration} {HotQCD}),\ }\href
  {\doibase 10.1103/PhysRevD.90.094503} {\bibfield  {journal} {\bibinfo
  {journal} {Phys. Rev.}\ }\textbf {\bibinfo {volume} {D90}},\ \bibinfo {pages}
  {094503} (\bibinfo {year} {2014})}\BibitemShut {NoStop}%
\bibitem [{\citenamefont {Bazavov}\ \emph {et~al.}(2012)\citenamefont {Bazavov}
  \emph {et~al.}}]{Bazavov:2012jq}%
  \BibitemOpen
  \bibfield  {author} {\bibinfo {author} {\bibfnamefont {A.}~\bibnamefont
  {Bazavov}} \emph {et~al.} (\bibinfo {collaboration} {HotQCD}),\ }\href
  {\doibase 10.1103/PhysRevD.86.034509} {\bibfield  {journal} {\bibinfo
  {journal} {Phys. Rev.}\ }\textbf {\bibinfo {volume} {D86}},\ \bibinfo {pages}
  {034509} (\bibinfo {year} {2012})}\BibitemShut {NoStop}%
\bibitem [{\citenamefont {Ding}\ \emph
  {et~al.}(2015{\natexlab{b}})\citenamefont {Ding}, \citenamefont {Mukherjee},
  \citenamefont {Ohno}, \citenamefont {Petreczky},\ and\ \citenamefont
  {Schadler}}]{Ding:2015fca}%
  \BibitemOpen
  \bibfield  {author} {\bibinfo {author} {\bibfnamefont {H.~T.}\ \bibnamefont
  {Ding}}, \bibinfo {author} {\bibfnamefont {S.}~\bibnamefont {Mukherjee}},
  \bibinfo {author} {\bibfnamefont {H.}~\bibnamefont {Ohno}}, \bibinfo {author}
  {\bibfnamefont {P.}~\bibnamefont {Petreczky}}, \ and\ \bibinfo {author}
  {\bibfnamefont {H.~P.}\ \bibnamefont {Schadler}},\ }\href {\doibase
  10.1103/PhysRevD.92.074043} {\bibfield  {journal} {\bibinfo  {journal} {Phys.
  Rev.}\ }\textbf {\bibinfo {volume} {D92}},\ \bibinfo {pages} {074043}
  (\bibinfo {year} {2015}{\natexlab{b}})}\BibitemShut {NoStop}%
\bibitem [{\citenamefont {Monnai}\ and\ \citenamefont
  {Schenke}(2016)}]{Monnai:2015sca}%
  \BibitemOpen
  \bibfield  {author} {\bibinfo {author} {\bibfnamefont {A.}~\bibnamefont
  {Monnai}}\ and\ \bibinfo {author} {\bibfnamefont {B.}~\bibnamefont
  {Schenke}},\ }\href {\doibase 10.1016/j.physletb.2015.11.063} {\bibfield
  {journal} {\bibinfo  {journal} {Phys. Lett.}\ }\textbf {\bibinfo {volume}
  {B752}},\ \bibinfo {pages} {317} (\bibinfo {year} {2016})}\BibitemShut
  {NoStop}%
\bibitem [{\citenamefont {Denicol}\ \emph {et~al.}(2016)\citenamefont
  {Denicol}, \citenamefont {Monnai},\ and\ \citenamefont
  {Schenke}}]{Denicol:2015nhu}%
  \BibitemOpen
  \bibfield  {author} {\bibinfo {author} {\bibfnamefont {G.}~\bibnamefont
  {Denicol}}, \bibinfo {author} {\bibfnamefont {A.}~\bibnamefont {Monnai}}, \
  and\ \bibinfo {author} {\bibfnamefont {B.}~\bibnamefont {Schenke}},\ }\href
  {\doibase 10.1103/PhysRevLett.116.212301} {\bibfield  {journal} {\bibinfo
  {journal} {Phys. Rev. Lett.}\ }\textbf {\bibinfo {volume} {116}},\ \bibinfo
  {pages} {212301} (\bibinfo {year} {2016})}\BibitemShut {NoStop}%
\bibitem [{\citenamefont {Buchel}(2008)}]{Buchel:2007mf}%
  \BibitemOpen
  \bibfield  {author} {\bibinfo {author} {\bibfnamefont {A.}~\bibnamefont
  {Buchel}},\ }\href {\doibase 10.1016/j.physletb.2008.03.069} {\bibfield
  {journal} {\bibinfo  {journal} {Phys. Lett.}\ }\textbf {\bibinfo {volume}
  {B663}},\ \bibinfo {pages} {286} (\bibinfo {year} {2008})}\BibitemShut
  {NoStop}%
\bibitem [{\citenamefont {Natsuume}\ and\ \citenamefont
  {Okamura}(2008)}]{Natsuume:2007ty}%
  \BibitemOpen
  \bibfield  {author} {\bibinfo {author} {\bibfnamefont {M.}~\bibnamefont
  {Natsuume}}\ and\ \bibinfo {author} {\bibfnamefont {T.}~\bibnamefont
  {Okamura}},\ }\href {\doibase 10.1103/PhysRevD.78.089902,
  10.1103/PhysRevD.77.066014} {\bibfield  {journal} {\bibinfo  {journal} {Phys.
  Rev.}\ }\textbf {\bibinfo {volume} {D77}},\ \bibinfo {pages} {066014}
  (\bibinfo {year} {2008})},\ \bibinfo {note} {[Erratum: Phys.
  Rev.D78,089902(2008)]}\BibitemShut {NoStop}%
\bibitem [{\citenamefont {Berdnikov}\ and\ \citenamefont
  {Rajagopal}(2000)}]{Berdnikov:1999ph}%
  \BibitemOpen
  \bibfield  {author} {\bibinfo {author} {\bibfnamefont {B.}~\bibnamefont
  {Berdnikov}}\ and\ \bibinfo {author} {\bibfnamefont {K.}~\bibnamefont
  {Rajagopal}},\ }\href {\doibase 10.1103/PhysRevD.61.105017} {\bibfield
  {journal} {\bibinfo  {journal} {Phys. Rev.}\ }\textbf {\bibinfo {volume}
  {D61}},\ \bibinfo {pages} {105017} (\bibinfo {year} {2000})}\BibitemShut
  {NoStop}%
\bibitem [{\citenamefont {Mukherjee}\ \emph {et~al.}(2015)\citenamefont
  {Mukherjee}, \citenamefont {Venugopalan},\ and\ \citenamefont
  {Yin}}]{Mukherjee:2015swa}%
  \BibitemOpen
  \bibfield  {author} {\bibinfo {author} {\bibfnamefont {S.}~\bibnamefont
  {Mukherjee}}, \bibinfo {author} {\bibfnamefont {R.}~\bibnamefont
  {Venugopalan}}, \ and\ \bibinfo {author} {\bibfnamefont {Y.}~\bibnamefont
  {Yin}},\ }\href {\doibase 10.1103/PhysRevC.92.034912} {\bibfield  {journal}
  {\bibinfo  {journal} {Phys. Rev.}\ }\textbf {\bibinfo {volume} {C92}},\
  \bibinfo {pages} {034912} (\bibinfo {year} {2015})}\BibitemShut {NoStop}%
\bibitem [{\citenamefont {Monnai}\ and\ \citenamefont
  {Hirano}(2009)}]{Monnai:2009ad}%
  \BibitemOpen
  \bibfield  {author} {\bibinfo {author} {\bibfnamefont {A.}~\bibnamefont
  {Monnai}}\ and\ \bibinfo {author} {\bibfnamefont {T.}~\bibnamefont
  {Hirano}},\ }\href {\doibase 10.1103/PhysRevC.80.054906} {\bibfield
  {journal} {\bibinfo  {journal} {Phys. Rev.}\ }\textbf {\bibinfo {volume}
  {C80}},\ \bibinfo {pages} {054906} (\bibinfo {year} {2009})}\BibitemShut
  {NoStop}%
\bibitem [{\citenamefont {Monnai}\ and\ \citenamefont
  {Hirano}(2010)}]{Monnai:2010qp}%
  \BibitemOpen
  \bibfield  {author} {\bibinfo {author} {\bibfnamefont {A.}~\bibnamefont
  {Monnai}}\ and\ \bibinfo {author} {\bibfnamefont {T.}~\bibnamefont
  {Hirano}},\ }\href {\doibase 10.1016/j.nuclphysa.2010.08.002} {\bibfield
  {journal} {\bibinfo  {journal} {Nucl. Phys.}\ }\textbf {\bibinfo {volume}
  {A847}},\ \bibinfo {pages} {283} (\bibinfo {year} {2010})}\BibitemShut
  {NoStop}%
\bibitem [{\citenamefont {Mukherjee}\ \emph {et~al.}(2016)\citenamefont
  {Mukherjee}, \citenamefont {Venugopalan},\ and\ \citenamefont
  {Yin}}]{Mukherjee:2016kyu}%
  \BibitemOpen
  \bibfield  {author} {\bibinfo {author} {\bibfnamefont {S.}~\bibnamefont
  {Mukherjee}}, \bibinfo {author} {\bibfnamefont {R.}~\bibnamefont
  {Venugopalan}}, \ and\ \bibinfo {author} {\bibfnamefont {Y.}~\bibnamefont
  {Yin}},\ }\href {\doibase 10.1103/PhysRevLett.117.222301} {\bibfield
  {journal} {\bibinfo  {journal} {Phys. Rev. Lett.}\ }\textbf {\bibinfo
  {volume} {117}},\ \bibinfo {pages} {222301} (\bibinfo {year}
  {2016})}\BibitemShut {NoStop}%
\bibitem [{\citenamefont {Karsch}\ \emph {et~al.}(2008)\citenamefont {Karsch},
  \citenamefont {Kharzeev},\ and\ \citenamefont {Tuchin}}]{Karsch:2007jc}%
  \BibitemOpen
  \bibfield  {author} {\bibinfo {author} {\bibfnamefont {F.}~\bibnamefont
  {Karsch}}, \bibinfo {author} {\bibfnamefont {D.}~\bibnamefont {Kharzeev}}, \
  and\ \bibinfo {author} {\bibfnamefont {K.}~\bibnamefont {Tuchin}},\ }\href
  {\doibase 10.1016/j.physletb.2008.01.080} {\bibfield  {journal} {\bibinfo
  {journal} {Phys. Lett.}\ }\textbf {\bibinfo {volume} {B663}},\ \bibinfo
  {pages} {217} (\bibinfo {year} {2008})}\BibitemShut {NoStop}%
\bibitem [{\citenamefont {Adamczyk}\ \emph {et~al.}(2014)\citenamefont
  {Adamczyk} \emph {et~al.}}]{Adamczyk:2014ipa}%
  \BibitemOpen
  \bibfield  {author} {\bibinfo {author} {\bibfnamefont {L.}~\bibnamefont
  {Adamczyk}} \emph {et~al.} (\bibinfo {collaboration} {STAR}),\ }\href
  {\doibase 10.1103/PhysRevLett.112.162301(2014),
  10.1103/PhysRevLett.112.162301} {\bibfield  {journal} {\bibinfo  {journal}
  {Phys. Rev. Lett.}\ }\textbf {\bibinfo {volume} {112}},\ \bibinfo {pages}
  {162301} (\bibinfo {year} {2014})}\BibitemShut {NoStop}%
\bibitem [{\citenamefont {Adamczyk}\ \emph {et~al.}(2016)\citenamefont
  {Adamczyk} \emph {et~al.}}]{Adamczyk:2016exq}%
  \BibitemOpen
  \bibfield  {author} {\bibinfo {author} {\bibfnamefont {L.}~\bibnamefont
  {Adamczyk}} \emph {et~al.} (\bibinfo {collaboration} {STAR}),\ }\href
  {\doibase 10.1103/PhysRevLett.116.112302} {\bibfield  {journal} {\bibinfo
  {journal} {Phys. Rev. Lett.}\ }\textbf {\bibinfo {volume} {116}},\ \bibinfo
  {pages} {112302} (\bibinfo {year} {2016})}\BibitemShut {NoStop}%
\bibitem [{\citenamefont {Song}\ and\ \citenamefont
  {Heinz}(2010)}]{Song:2009rh}%
  \BibitemOpen
  \bibfield  {author} {\bibinfo {author} {\bibfnamefont {H.}~\bibnamefont
  {Song}}\ and\ \bibinfo {author} {\bibfnamefont {U.~W.}\ \bibnamefont
  {Heinz}},\ }\href {\doibase 10.1103/PhysRevC.81.024905} {\bibfield  {journal}
  {\bibinfo  {journal} {Phys. Rev.}\ }\textbf {\bibinfo {volume} {C81}},\
  \bibinfo {pages} {024905} (\bibinfo {year} {2010})}\BibitemShut {NoStop}%
\bibitem [{\citenamefont {Noronha-Hostler}\ \emph {et~al.}(2013)\citenamefont
  {Noronha-Hostler}, \citenamefont {Denicol}, \citenamefont {Noronha},
  \citenamefont {Andrade},\ and\ \citenamefont
  {Grassi}}]{Noronha-Hostler:2013gga}%
  \BibitemOpen
  \bibfield  {author} {\bibinfo {author} {\bibfnamefont {J.}~\bibnamefont
  {Noronha-Hostler}}, \bibinfo {author} {\bibfnamefont {G.~S.}\ \bibnamefont
  {Denicol}}, \bibinfo {author} {\bibfnamefont {J.}~\bibnamefont {Noronha}},
  \bibinfo {author} {\bibfnamefont {R.~P.~G.}\ \bibnamefont {Andrade}}, \ and\
  \bibinfo {author} {\bibfnamefont {F.}~\bibnamefont {Grassi}},\ }\href
  {\doibase 10.1103/PhysRevC.88.044916} {\bibfield  {journal} {\bibinfo
  {journal} {Phys. Rev.}\ }\textbf {\bibinfo {volume} {C88}},\ \bibinfo {pages}
  {044916} (\bibinfo {year} {2013})}\BibitemShut {NoStop}%
\bibitem [{\citenamefont {Noronha-Hostler}\ \emph {et~al.}(2014)\citenamefont
  {Noronha-Hostler}, \citenamefont {Noronha},\ and\ \citenamefont
  {Grassi}}]{Noronha-Hostler:2014dqa}%
  \BibitemOpen
  \bibfield  {author} {\bibinfo {author} {\bibfnamefont {J.}~\bibnamefont
  {Noronha-Hostler}}, \bibinfo {author} {\bibfnamefont {J.}~\bibnamefont
  {Noronha}}, \ and\ \bibinfo {author} {\bibfnamefont {F.}~\bibnamefont
  {Grassi}},\ }\href {\doibase 10.1103/PhysRevC.90.034907} {\bibfield
  {journal} {\bibinfo  {journal} {Phys. Rev.}\ }\textbf {\bibinfo {volume}
  {C90}},\ \bibinfo {pages} {034907} (\bibinfo {year} {2014})}\BibitemShut
  {NoStop}%
\bibitem [{\citenamefont {Ryu}\ \emph {et~al.}(2015)\citenamefont {Ryu},
  \citenamefont {Paquet}, \citenamefont {Shen}, \citenamefont {Denicol},
  \citenamefont {Schenke}, \citenamefont {Jeon},\ and\ \citenamefont
  {Gale}}]{Ryu:2015vwa}%
  \BibitemOpen
  \bibfield  {author} {\bibinfo {author} {\bibfnamefont {S.}~\bibnamefont
  {Ryu}}, \bibinfo {author} {\bibfnamefont {J.~F.}\ \bibnamefont {Paquet}},
  \bibinfo {author} {\bibfnamefont {C.}~\bibnamefont {Shen}}, \bibinfo {author}
  {\bibfnamefont {G.~S.}\ \bibnamefont {Denicol}}, \bibinfo {author}
  {\bibfnamefont {B.}~\bibnamefont {Schenke}}, \bibinfo {author} {\bibfnamefont
  {S.}~\bibnamefont {Jeon}}, \ and\ \bibinfo {author} {\bibfnamefont
  {C.}~\bibnamefont {Gale}},\ }\href {\doibase 10.1103/PhysRevLett.115.132301}
  {\bibfield  {journal} {\bibinfo  {journal} {Phys. Rev. Lett.}\ }\textbf
  {\bibinfo {volume} {115}},\ \bibinfo {pages} {132301} (\bibinfo {year}
  {2015})}\BibitemShut {NoStop}%
\bibitem [{\citenamefont {Torrieri}\ \emph {et~al.}(2008)\citenamefont
  {Torrieri}, \citenamefont {Tomasik},\ and\ \citenamefont
  {Mishustin}}]{Torrieri:2007fb}%
  \BibitemOpen
  \bibfield  {author} {\bibinfo {author} {\bibfnamefont {G.}~\bibnamefont
  {Torrieri}}, \bibinfo {author} {\bibfnamefont {B.}~\bibnamefont {Tomasik}}, \
  and\ \bibinfo {author} {\bibfnamefont {I.}~\bibnamefont {Mishustin}},\ }\href
  {\doibase 10.1103/PhysRevC.77.034903} {\bibfield  {journal} {\bibinfo
  {journal} {Phys. Rev.}\ }\textbf {\bibinfo {volume} {C77}},\ \bibinfo {pages}
  {034903} (\bibinfo {year} {2008})}\BibitemShut {NoStop}%
\bibitem [{\citenamefont {Torrieri}\ and\ \citenamefont
  {Mishustin}(2008)}]{Torrieri:2008ip}%
  \BibitemOpen
  \bibfield  {author} {\bibinfo {author} {\bibfnamefont {G.}~\bibnamefont
  {Torrieri}}\ and\ \bibinfo {author} {\bibfnamefont {I.}~\bibnamefont
  {Mishustin}},\ }\href {\doibase 10.1103/PhysRevC.78.021901} {\bibfield
  {journal} {\bibinfo  {journal} {Phys. Rev.}\ }\textbf {\bibinfo {volume}
  {C78}},\ \bibinfo {pages} {021901} (\bibinfo {year} {2008})}\BibitemShut
  {NoStop}%
\bibitem [{\citenamefont {Rajagopal}\ and\ \citenamefont
  {Tripuraneni}(2010)}]{Rajagopal:2009yw}%
  \BibitemOpen
  \bibfield  {author} {\bibinfo {author} {\bibfnamefont {K.}~\bibnamefont
  {Rajagopal}}\ and\ \bibinfo {author} {\bibfnamefont {N.}~\bibnamefont
  {Tripuraneni}},\ }\href {\doibase 10.1007/JHEP03(2010)018} {\bibfield
  {journal} {\bibinfo  {journal} {JHEP}\ }\textbf {\bibinfo {volume} {03}},\
  \bibinfo {pages} {018} (\bibinfo {year} {2010})}\BibitemShut {NoStop}%
\bibitem [{\citenamefont {Habich}\ and\ \citenamefont
  {Romatschke}(2014)}]{Habich:2014tpa}%
  \BibitemOpen
  \bibfield  {author} {\bibinfo {author} {\bibfnamefont {M.}~\bibnamefont
  {Habich}}\ and\ \bibinfo {author} {\bibfnamefont {P.}~\bibnamefont
  {Romatschke}},\ }\href {\doibase 10.1007/JHEP12(2014)054} {\bibfield
  {journal} {\bibinfo  {journal} {JHEP}\ }\textbf {\bibinfo {volume} {12}},\
  \bibinfo {pages} {054} (\bibinfo {year} {2014})}\BibitemShut {NoStop}%
\bibitem [{\citenamefont {Schofield}\ \emph {et~al.}(1969)\citenamefont
  {Schofield}, \citenamefont {Litster},\ and\ \citenamefont
  {Ho}}]{PhysRevLett.23.1098}%
  \BibitemOpen
  \bibfield  {author} {\bibinfo {author} {\bibfnamefont {P.}~\bibnamefont
  {Schofield}}, \bibinfo {author} {\bibfnamefont {J.~D.}\ \bibnamefont
  {Litster}}, \ and\ \bibinfo {author} {\bibfnamefont {J.~T.}\ \bibnamefont
  {Ho}},\ }\href {\doibase 10.1103/PhysRevLett.23.1098} {\bibfield  {journal}
  {\bibinfo  {journal} {Phys. Rev. Lett.}\ }\textbf {\bibinfo {volume} {23}},\
  \bibinfo {pages} {1098} (\bibinfo {year} {1969})}\BibitemShut {NoStop}%
\bibitem [{\citenamefont {Zinn-Justin}(2001)}]{ZinnJustin:1999bf}%
  \BibitemOpen
  \bibfield  {author} {\bibinfo {author} {\bibfnamefont {J.}~\bibnamefont
  {Zinn-Justin}},\ }\href {\doibase 10.1016/S0370-1573(00)00126-5} {\bibfield
  {journal} {\bibinfo  {journal} {Phys. Rept.}\ }\textbf {\bibinfo {volume}
  {344}},\ \bibinfo {pages} {159} (\bibinfo {year} {2001})}\BibitemShut
  {NoStop}%
\bibitem [{\citenamefont {Stephanov}(2011)}]{Stephanov:2011pb}%
  \BibitemOpen
  \bibfield  {author} {\bibinfo {author} {\bibfnamefont {M.}~\bibnamefont
  {Stephanov}},\ }\href {\doibase 10.1103/PhysRevLett.107.052301} {\bibfield
  {journal} {\bibinfo  {journal} {Phys. Rev. Lett.}\ }\textbf {\bibinfo
  {volume} {107}},\ \bibinfo {pages} {052301} (\bibinfo {year}
  {2011})}\BibitemShut {NoStop}%
\end{thebibliography}%

\begin{appendix}
\section{Parametrization of correlation in Ising variables
\label{sec:IS}
}

For completeness, in this section,
we explain the parameterization of the critical correlation length
$\xi(r,h)$ in the critical regime in terms of the Ising
variables $r$ and $h$ used in this paper.
For this purpose,
we only need to know the equilibrium magnetization $M(r,h)$ as
$\xi$ can be computed by taking derivatives of
$M(r,h)$ with respect to $h$ at fixed $r$,
\be
\label{eq:M_kappa}
\xi^{2} =\frac{1}{ H_{0} }
\(\frac{\pd M (r,h)}{\pd h}\)_{r}\, .
\ee
Here $H_{0}$ is a dimensionful parameter (of mass dimension $3$) which relates reduced
magnetic field $h$ to the un-reduced magnetic field.
As explained earlier (c.f.~footnote 1),
we will take critical exponent $\eta=0$. 
Therefore R.H.S of \eqref{eq:M_kappa}, which is nothing but magnetic susceptibility $\chi_{M}$ in Ising model, is proportional to $\xi^{2}$.
This relation have been widely used in the previous studies (see for example Ref.~\cite{Stephanov:2008qz}).

To parametrize $M^{\equ}(r,h)$,
we use  the linear parametric model \cite{PhysRevLett.23.1098,ZinnJustin:1999bf}.
In this parametrization,
one introduces two new variables $R,\theta$ which are related to (dimensionless) Ising
variable $r, h$ as
\be
\label{eq:rh_Rtheta}
r(R,\theta)= R(1-\theta ^2)\, ,
\qquad
h(R,\theta)= \Delta h\, R^{\b \d}\,\tilde{h}(\theta)\, ,
\ee
Following Ref.~\cite{Stephanov:2011pb},
we will use
\be
\label{eq:th}
\tilde{h}(\theta) = 3\theta \[1-\(\frac{(\delta-1)(1-2\beta)}{(\delta-3) }\)\theta^2\]\, .
\ee
Here $\b,\d$ are standard critical exponents and we will use the values
obtained from mean field theory, $\b=1/3,\delta = 5$. 
In these $R,\theta$ variables, $\theta=0$ corresponds to the crossover
line and $|\theta|=\sqrt{3/2}$ 
corresponds to the coexistence (first order transition) line. The  equilibrium ``magnetization'' $M^{\equ}_{0}(r,h)$(or $\s_{0}$)
is given by
\be
\label{eq:M_Rtheta}
M^{\equ}(R,\theta)
= M_{0}R^{\b}\theta\, ,
\ee
where $M_{0}$ sets the scale of ``magnetization''. 
The parametrization introduced describes the equation of state with a precision
sufficient for our purpose.

We now compute $\k^{\equ}_{n}$ using \eq\eqref{eq:M_kappa} and \eq\eqref{eq:M_Rtheta}.
Explicitly, we have
\begin{eqnarray}
\label{eq:kappa_rh}
\xi^{2}
&=&\frac{M_{0}}{H_{0}}\frac{1}{R^{4/3}(3+2\theta^2)}\, .
\end{eqnarray}
One could then determine $\xi(r,h)$ consequently from \eqref{eq:kappa_rh} using \eqref{eq:rh_Rtheta}. 

\end{appendix}

\newpage

Notice: This manuscript has been co-authored by employees of Brookhaven Science Associates, LLC under Contract No. DE-SC0012704 with the U.S. Department of Energy. The publisher by accepting the manuscript for publication acknowledges that the United States Government retains a non-exclusive, paid-up, irrevocable, world-wide license to publish or reproduce the published form of this manuscript, or allow others to do so, for United States Government purposes. This preprint is intended for publication in a journal or proceedings. Since changes may be made before publication, it may not be cited or reproduced without the authors permission.

DISCLAIMER: This report was prepared as an account of work sponsored by an agency of the United States Government. Neither the United States Government nor any agency thereof, nor any of their employees, nor any of their contractors, subcontractors, or their employees, makes any warranty, express or implied, or assumes any legal liability or responsibility for the accuracy, completeness, or any third partys use or the results of such use of any information, apparatus, product, or process disclosed, or represents that its use would not infringe privately owned rights. Reference herein to any specific commercial product, process, or service by trade name, trademark, manufacturer, or otherwise, does not necessarily constitute or imply its endorsement, recommendation, or favoring by the United States Government or any agency thereof or its contractors or subcontractors. The views and opinions of authors expressed herein do not necessarily state or reflect those of the United States Government or any agency thereof.

\end{document}